\documentclass[aps,prd,preprintnumbers,groupedaddress,nofootinbib,showpacs]{revtex4}

\pdfoutput=1

\usepackage{graphicx}
\usepackage{latexsym}
\usepackage{amsfonts}
\usepackage{amssymb}
\usepackage{amsmath}
\usepackage{slashed}
\usepackage{array}
\usepackage{feynmp}
\usepackage{url}
\begin{document}
\title{Using Effective Operators to Understand CoGeNT and CDMS-Si}

\author{Matthew R.~Buckley$^{1}$}
\affiliation{$^1$Center for Particle Astrophysics, Fermi National Accelerator Laboratory, Batavia, IL 60510, USA}
\preprint{FERMILAB-PUB-}
\date{\today}

\begin{abstract}
Several direct detection experiments have reported positive signals consistent with a dark matter particle with a mass of approximately $7-9$~GeV and a spin independent scattering cross section of $2.5-4.8 \times 10^{-41}$~cm$^2$. These results do not rise to the level of discovery, but assuming that they are due to dark matter, some questions about the underlying physics can already be addressed. In this paper, I apply the effective operator formalism for dark matter-Standard Model interactions to the results of the CoGeNT and CDMS silicon target experiments. I demonstrate that only one set of flavor-blind effective operators between dark matter can quarks can be consistent with the reported results in all energy regimes of interest, namely thermal freeze-out, nuclear scattering, indirect detection, and TeV-scale colliders. This set of operators implies large couplings of dark matter with heavy quarks. The alternative implies either that the new physics has non-trivial flavor structure, that the effective formalism is not applicable and so contains new states in the spectrum accessible at the LHC, or has large annihilation channels (possibly via effective operators) into non-colored Standard Model particles.
\end{abstract}


\maketitle

\section{Introduction \label{sec:intro}}

A growing number of dark matter direct detection experiments have reported anomalous events that can be interpreted as signals of dark matter. Assuming elastic scattering, these signals suggest dark matter with a mass of $7-9$~GeV, rather than ${\cal O}(100~\mbox{GeV})$ as expected in the most straightforward interpretations of the ``WIMP miracle." 
Three experiments -- DAMA/LIBRA \cite{Bernabei:2010mq}, CoGeNT \cite{Aalseth:2010vx,Aalseth:2011wp}, and the silicon data from CDMS-II \cite{Agnese:2013rvf} (CDMS-Si) -- each report events broadly compatible with the same signal of ``light'' dark matter. A fourth experiment, CRESST-II \cite{Angloher:2011uu}, reported an excess that appears to be consistent of somewhat heavier dark matter, though a possible issue with backgrounds caused by surface scattering was identified \cite{Kuzniak:2012zm}, which may have a large effect on the results. 

Experiments based on liquid xenon should have sensitivity to dark matter in the range of masses and cross sections of interest. Neither XENON100 \cite{Aprile:2012nq} or the XENON10 S2-only analysis \cite{Angle:2011th} have seen events compatible with the claimed signal. It is possible that a non-Maxwell-Boltzmann velocity distribution \cite{Mao:2013nda} or non-trivial particle physics (for example, exothermic or isospin-violating dark matter \cite{Kurylov:2003ra,Giuliani:2005my,Frandsen:2013cna,DelNobile:2013cta}) could alleviate this tension. The response of xenon to low-energy recoils is also a matter of some debate \cite{Collar:2011wq,Horn:2011wz,Manalaysay:2010mb,Manzur:2009hp,Plante:2011hw,Hooper:2013cwa}, and could result in a significant reduction of sensitivity from the xenon-based experiments. Additionally, though the prospective signals from DAMA/LIBRA, CDMS-Si, and CoGeNT are usually interpreted in terms of spin-independent interactions between dark matter and the target nuclei, a spin-dependent interpretation is not firmly ruled out \cite{Buckley:2013gjo}. In particular, a spin-dependent interaction with neutrons is in approximately the same amount of tension with the XENON100 results \cite{Aprile:2013doa} as a spin-independent interaction, though such a dark matter candidate would also be in some tension with PICASSO results \cite{Archambault:2012}.

Much more data is required before the CoGeNT, DAMA/LIBRA, and CDMS-Si results can be seen as evidence of dark matter. However, even at this early date we can begin to ask questions of the underlying particle physics mediating the direct detection interaction, assuming that these signals are due to dark matter. Though there are many possible theories that could result in a dark matter-nucleon cross section of the appropriate size, a useful parameterization is in terms of effective operators (see {\it e.g.}~\cite{Goodman:2010ku}). Here, the dark matter-quark (or dark matter-gluon) coupling is written in terms of higher-dimension operators suppressed by some scale $\Lambda$. The validity of this approach assumes that the physics that mediates the direct detection is at a high scale compared to the energy transfer involved in the relevant process. If this effective operator formalism continues to hold true at energies equivalent to the rest mass of the dark matter and the center of mass energy for collisions at the LHC, then the same effective operator will result in both a thermal cross section in the early Universe, indirect detection of annihilations occurring today in the Universe, and the production of dark matter in the collider experiments \cite{Beltran:2010ww,Goodman:2010ku,Bai:2012xg,Carpenter:2012rg,Fox:2011pm}.

The implications of the CoGeNT and CDMS-Si results for effective operators bounds at colliders has been considered recently in Ref.~\cite{Cheung:2013pfa}, which identified which operators could give the observed signals without violating constraints from collider physics. In this paper, I consider direct detection, indirect detection constraints from the Fermi Gamma-Ray Telescope's observations of dwarf galaxies \cite{GeringerSameth:2011iw,Ackermann:2011wa}, thermal relic abundance, and collider constraints on effective operators in light of the ${\cal O}(10~\mbox{GeV})$ dark matter suggested by the DAMA/LIBRA, CoGeNT, and CDMS-Si results. I work with the possible set of operators coupling quarks and gluons to scalar or fermionic dark matter.\footnote{The effective operator formalism's application to vector dark matter has not been as widely considered in the community. While an interesting possibility, it is beyond the scope of this paper.
}
I first consider whether a single minimally flavor-violating operator could explain the direct detection signals while also providing a sufficiently large cross section to allow for dark matter to be thermally produced in the early Universe with the correct relic abundance. Considering both spin-independent and spin-dependent interactions, only a single operator can perform this task without violating collider or dwarf galaxy bounds. For the remaining operators that could induce a direct detection signal, there are no additional operators that can be added to the theory which do not induce a direct detection signal and also provide a sufficiently large early Universe annihilation cross section while avoiding present collider constraints. 

The single exception is a real or complex scalar dark matter particle coupling through scalar operators. Such operators have comparatively weak bounds, due to dark matter/nucleon couplings that are proportional to the light quark masses. However, such a coupling structure suggests that the dark matter would have a large interaction with bottom and top quarks. Searches in these channels have been suggested \cite{Lin:2013sca}, but do not yet reach the level of sensitivity necessary to discover or exclude operators of the size suggested by the direct detection results. Interestingly, the indicated energy scale lies close to the regime where the effective operator formalism may not be applicable \cite{Busoni:2013lha,Profumo:2013hqa}, which could allow for larger (model-dependent) signals at the LHC. 
 
This set of arguments implies that the underlying physics must fall into one of the four following categories. Either 
\begin{itemize}
\item the dark matter is a scalar interacting with the Standard Model quarks with scalar operators that have couplings proportional to the quark mass, or

\item the dark matter-quark interactions are not described by an effective operator at colliders, direct or indirect detection, and/or the early Universe, or

\item the dark matter has an annihilation channel into something other than quarks described by an operator distinct from the effective operator to quarks, or

\item there is non-trivial flavor structure in the dark matter-quark couplings.
\end{itemize}
All of these options are highly suggestive of a dark matter sector with new physics at scales that may be accessible in either the LHC or future colliders. If dark matter does couple to quark mass, then there is a possibility that dedicated collider searches in $b$- and $t$-rich channels could see evidence of the new physics. If interactions with quarks cannot  be written as an effective operator, then this requires some new light states through which the interaction can proceed, states that must couple to quarks. If there are other large interactions providing the necessary annihilation in the early Universe, this suggests either some hidden sector, or large couplings with the leptons in the Standard Model. Couplings that are not minimally flavor violating yet do have not made themselves known through precision flavor tests would tie the problem of dark matter into the flavor problem. 

In the next section, I review the effective operator formalism used throughout the remainder of the paper. I adapt the notation used in Ref.~\cite{Goodman:2010ku}, which has become a standardized way for experimental collaborations to report their search results \cite{Chatrchyan:2012tea,Chatrchyan:2012me,Aad:2012awa,Aad:2012fw,ATLAS:2012ky,Zhou:2013fla}. I show the bounds on the effective operator scales derived from direct searches at colliders, as well as those extrapolated by theorists assuming heavy flavor interactions. In Section~\ref{sec:dd}, I consider the effective operators that can produce the spin-independent or spin-dependent signals in the direct detection experiments, and show that none of these operators can simultaneously provide the observed rates of events while also realizing a sufficiently large thermal cross section and avoiding the collider bounds. In Section~\ref{sec:others}, I review the remaining operators that do not result in direct detection signals, but could contribute to annihilation in the early Universe. I will show that the required scales for these operators are all ruled out by collider constraints, assuming minimal flavor violation, or lie in a region where the effective operator approximation breaks down. For readers looking for a summary of the results, the constraints on the scale $\Lambda$ for each operator under consideration (see Table~\ref{tab:listofoperators}) can be found in Figs.~\ref{fig:signal} and \ref{fig:signal2} (for operators resulting in direct detection signals) and Figs.~\ref{fig:others} and \ref{fig:others2} (for operators that do not induce such signals). 

\section{Effective Operators \label{sec:effective}}

At energies much lower than the mass of a particle, that particle can be integrated out of the theory. This collapses the finite-range interactions of lighter fields mediated by the heavy particle down to a four-point contact interaction. In the low-energy effective Lagrangian this term has dimension $>4$ (unless the effective operator connects four scalars, which is not of interest here, as direct detection requires interaction with Standard Model quarks or gluons), and so is suppressed by some mass scale $\Lambda$. 

Integrating out the heavy physics of course makes it difficult to reconstruct the full theory, but imprints of the high energy interaction remain in the Lorentz structure of the effective operator. For example, the chiral nature of the weak interaction is visible in the left- and right-handed projection operators in the Fermi theory. Furthermore, while the scale $\Lambda$ can be interpreted as the mass scale of the mediating particle, it is actually a combination of couplings and masses. Again using the familiar language of the effective theory of the weak interaction, the Fermi constant is given by $G_F = \sqrt{2} g^2/8m_W^2$. Writing this in the language I will use throughout the paper, the Fermi interaction is a four-fermion contact interaction suppressed by $\Lambda^{-2}$, with $\Lambda = (8/\sqrt{2})^{1/2} m_W/g$. Notice that $\Lambda \gtrsim M$, where $M$ is the mass of the heavy particle. Equality is achieved when the coupling is non-perturbative.

Assuming that dark matter is some massive particle which was produced thermally in the early Universe, we can postulate that the dark matter coupling to Standard Model fields is mediated by some effective operator suppressed by a scale $\Lambda$. Without any further experimental input, there are many possible operators that could be relevant. Indeed, as we do not know the spin assignment of dark matter, the full list of effective operators must include both boson and fermion dark matter particles. 

In Table~\ref{tab:listofoperators}, I reproduce the list of effective operators from Ref.~\cite{Goodman:2010ku} which connect dark matter (denoted $\chi$ for both scalars and fermions) to Standard Model fermions ($f$) and gluons up to dimension-six, assuming either spin-0 or spin-$\tfrac{1}{2}$ dark matter. The naming convention has become standard, and the prefix ``D," ``C,'' or ``R'' refers to Dirac fermion, complex scalar, or real scalar dark matter, respectively. I will also consider Majorana dark matter, which will use the same set of operators as the Dirac fermions. Apart from additional numerical factors, this case differs from Dirac fermions in that several operators (D5, D7, D9, and D10) which allow annihilation and scattering for Dirac fermions vanish for Majorana dark matter.

\begin{table}[h]
\begin{tabular}{|c|c|c|c|c|}
\hline
Name & Operator & Coefficient & DD \\ \hline \hline
D1 & $[\bar{\chi}\chi][\bar{f}f]$ & $m_f\Lambda^{-3}$ & SI \\ \hline
D2 &	 $[\bar{\chi}\gamma^5\chi][\bar{f}f]$ & $im_f \Lambda^{-3}$ & -- \\ \hline
D3 & $[\bar{\chi}\chi][\bar{f} \gamma^5 f]$ &$im_f \Lambda^{-3}$ & -- \\ \hline
D4 & $[\bar{\chi}\gamma^5 \chi][\bar{f} \gamma^5 f]$ & $m_f \Lambda^{-3}$ & -- \\ \hline
D5 & $[\bar{\chi}\gamma^\mu \chi][\bar{f} \gamma_\mu f]$ & $\Lambda^{-2}$ & SI \\ \hline
D6 & $[\bar{\chi}\gamma^\mu\gamma^5 \chi][\bar{f} \gamma_\mu  f]$ & $\Lambda^{-2} $ & -- \\ \hline
D7 & $[\bar{\chi}\gamma^\mu \chi][\bar{f} \gamma_\mu\gamma^5  f]$ & $\Lambda^{-2}$ & -- \\ \hline
D8 & $[\bar{\chi}\gamma^\mu \gamma^5 \chi][\bar{f} \gamma_\mu\gamma^5  f]$ & $\Lambda^{-2}$ & SD\\ \hline
D9 & $[\bar{\chi}\sigma^{\mu\nu} \chi] [\bar{f} \sigma_{\mu\nu} f]$ & $\Lambda^{-2}$ & SD \\ \hline
D10 & $[\bar{\chi}\sigma^{\mu\nu}\gamma^5 \chi] [\bar{f} \sigma_{\mu\nu} f]$ & $i\Lambda^{-2}$ & -- \\ \hline
D11 & $[\bar{\chi}\chi] [G_{\mu\nu}G^{\mu\nu}]$ & $\alpha_S\Lambda^{-3}$ & SI \\ \hline
D12 & $[\bar{\chi}\gamma^5 \chi] [G_{\mu\nu}G^{\mu\nu}]$ & $i\alpha_S\Lambda^{-3}$ & -- \\ \hline
D13 & $[\bar{\chi}\chi] [G_{\mu\nu}\tilde{G}^{\mu\nu}]$ & $i\alpha_S\Lambda^{-3}$ & -- \\ \hline
D14 & $[\bar{\chi}\gamma^5\chi] [G_{\mu\nu}\tilde{G}^{\mu\nu}]$ & $\alpha_S\Lambda^{-3} $ & -- \\ \hline
\end{tabular}
\begin{tabular}{|c|c|c|c|}
\hline
Name & Operator & Coefficient & DD \\ \hline \hline
C1 & $[\chi^*\chi][\bar{f}f]$ & $m_f\Lambda^{-2}$ & SI \\ \hline
C2 & $[\chi^*\chi][\bar{f}\gamma^5f]$ & $im_f\Lambda^{-2}$ & -- \\ \hline
C3 & $[\chi^*\partial_\mu \chi][\bar{f}\gamma^\mu f]$ & $\Lambda^{-2}$ & SI \\ \hline
C4 & $[\chi^*\partial_\mu \chi][\bar{f}\gamma^\mu \gamma^5 f]$ & $\Lambda^{-2}$ & -- \\ \hline
C5 & $[\chi^* \chi][G_{\mu\nu}G^{\mu\nu}]$ & $\alpha_S\Lambda^{-2}$ &  SI \\ \hline
C6 & $[\chi^*\chi][G_{\mu\nu}\tilde{G}^{\mu\nu}]$ & $i\alpha_S\Lambda^{-2}$ & -- \\ \hline \hline
R1 & $[\chi \chi][\bar{f}f]$ & $m_f\Lambda^{-2}$ & SI \\ \hline
R2 & $[\chi \chi][\bar{f}\gamma^5f]$ & $im_f\Lambda^{-2}$ & -- \\ \hline
R3 & $[\chi \chi][G_{\mu\nu}G^{\mu\nu}]$ & $\alpha_S\Lambda^{-2}$ & SI  \\ \hline
R4 & $[\chi\chi][G_{\mu\nu}\tilde{G}^{\mu\nu}]$ & $i\alpha_S\Lambda^{-2}$ & -- \\ \hline
\end{tabular}
\caption{List of effective operators from Ref.~\cite{Goodman:2010ku}, along with assumed normalization and the type of elastic direct detection (DD) induced. Direct detection can be either spin-independent (SI) or spin-dependent (SD). Operators with neither a SI or SD designation induce elastic scattering that is suppressed by powers of dark matter velocity $v$ or momentum transfer $q$ \cite{Kumar:2013iva}. Operators for Dirac fermion dark matter are denoted by ``D,'' complex scalar dark matter by ``C,'' and real scalars by ``R.'' Note that the normalization of some of the operator coefficients  differ from those in Ref.~\cite{Goodman:2010ku}. \label{tab:listofoperators}}
\end{table}

To prevent dangerous flavor-violating effects, I make the standard assumption that scalar and pseudoscalar operators (of the form $\bar{f}f$ or $\bar{f}\gamma^5f$) are proportional to the Standard Model fermion mass (minimal flavor violation \cite{DAmbrosio:2002ex}). 
This results in operators suppressed by an additional power of the high scale $\Lambda$, compared to the na\"{i}ve dimension counting. Beyond this, I assume that the operators are flavor-conserving. Where applicable, I will mention the effects of relaxing this assumption. 

Operators connecting dark matter to gluons have an explicit strong-coupling constant $\alpha_S$ factored out. This reflects the assumption that, since dark matter is an $SU(3)_C$ singlet, couplings to gluons should proceed through a loop of colored particles. Note that the normalization for some operators differs from that chosen by Ref.~\cite{Goodman:2010ku}. I take these differences into account when translating bounds from other works.

In this paper, I am assuming that the light dark matter signals seen in direct detection are being generated by an effective operator from Table~\ref{tab:listofoperators}. As we do not yet know with certainty whether the putative signal being observed at the experiments is due to a spin dependent or spin independent coupling between the nuclei and dark matter, we must consider both options. 

In addition to requiring a large enough coupling ({\it i.e.}~a small enough $\Lambda$) to give the observed events in direct detection, if dark matter's only interaction with other particles is through an effective operator, then assuming dark matter is a thermal relic of the early Universe, some operator must have a low enough $\Lambda$ so that efficient annihilation into quarks, leptons, or gauge bosons could have occurred. Of the operators in Table~\ref{tab:listofoperators}, none will interfere with each other in annihilation processes \cite{Kumar:2013iva}. Therefore, we can consider the early Universe thermal cross section of each in turn. Expanding the thermal cross section in the temperature to ${\cal O}(T)$ for each operator, I list the results in Appendix~\ref{app:thermal}. 

For a cross section $\langle \sigma v\rangle = a+bT$, the combination that appears in the calculation of the thermal relic density is $a+\frac{1}{2x_f} b$, where $x_f \equiv T_f/m_\chi\sim 20$ is the ratio of the freeze-out temperature to the mass of the dark matter. To obtain the dark matter relic density $\Omega h^2 =0.119$ \cite{Ade:2013zuv}
the combination $a+\frac{1}{2 x_f}b$ must be $2 \times 10^{-9}$~GeV$^{-2}$ for Majorana fermions and real scalars.\footnote{For a recent review of freeze-out calculations, see Ref.~\cite{Steigman:2012nb}.} For Dirac fermions and complex scalars, the effective cross section must be twice as large, as the dark matter is composed of both particles and antiparticles, each with half the observed relic density. From this, I derive the required scale $\Lambda$ for each operator assuming that each operator is uniquely responsible for providing the thermal cross section for dark matter, and that annihilation proceeds into all kinematically accessible Standard Model final states. For the operators coupling dark matter to fermions, I include annihilation into leptons in this calculation, as this results in a conservative estimation for the required scale $\Lambda$. The required values of $\Lambda$ for each operator will be discussed in more detail in Sections~\ref{sec:dd} and Section~\ref{sec:others}.

This thermal cross section can also be apparent in the Universe today, through the indirect detection of dark matter annihilation. This requires the thermal cross section to have a part that is not velocity suppressed, as the dark matter in the present Universe is moving much slower than the speed of light ($T \ll m_\chi$). The most stringent bounds on indirect detection in the channels and mass range of interest come from the Fermi Gamma-Ray Space Telescope's observations of dwarf galaxies local to the Milky Way \cite{GeringerSameth:2011iw,Ackermann:2011wa}. Depending on the annihilation channel, these observations place upper limits on the annihilation rate of dark matter, and thus the velocity averaged cross section which are competitive with the canonical cross section for thermal freeze-out. In this paper, I use the upper limits on the cross section in the $\bar{b}b$ annihilation channel from Ref.~\cite{Ackermann:2011wa}. Other, more aggressive limits exist \cite{Hooper:2012sr}, and using additional channels would also place more stringent bounds on the cross sections and therefore the suppression scales $\Lambda$ in the effective formalism. However, the indirect bounds will be of most interest for operators coupling through Standard Model fermion mass ({\it i.e.}~operators such as C1 and R1), and so concentrating on the $b$-quark annihilation channel is both relevant and conservative. For dark matter that is not its own antiparticle, a correction factor of 2 is applied to the bounds of Ref.~\cite{Ackermann:2011wa}.

Finally, if dark matter couples to quarks or gluons, then it will be pair-produced in colliders. The dark matter itself will be invisible, but will typically be produced in association with a high $p_T$ jet \cite{Beltran:2010ww,Goodman:2010yf,Goodman:2010ku,Fox:2011pm,Rajaraman:2011wf,Fox:2012ee}, photon \cite{Fox:2011pm,Fox:2011fx}, or $W/Z$ \cite{Bai:2012xg,Aad:2012awa,Carpenter:2012rg}, allowing for searches looking for unbalanced events with large missing $p_T$. Assuming that the operator formalism continues to hold at LHC center-of-mass energies, then the lack of signal allows the experimental collaborations to place an lower bound on $\Lambda$ for each operator. In this paper, I will adopt the most recent combined analysis, Ref.~\cite{Zhou:2013fla}, using all visible final states. The resulting bounds for each operator (dubbed ``mono-everything'') are listed Table~\ref{tab:collider}. The bounds on Majorana and real scalar dark matter are weaker than those corresponding to Dirac and complex dark matter operators by a factor of $2^{-1/2n}$ where $n = 2,3$ is the power of the operator's suppression scale $\Lambda$. This is because of a relative factor of $\tfrac{1}{2}$ in the production cross section for identical final states. 

\begin{table}[ht]
\begin{tabular}{|c|c|}
\hline
Name & Bound~(GeV) \\ \hline \hline
D1-D4 & 34 \\
(Majorana) & 30 \\ \hline
D5 & 795 \\ \hline
D6 & 791 \\ \hline
D7 & 812 \\ \hline
D8 & 811 \\ \hline
D9 & 1331 \\
(Majorana) & 1119 \\ \hline
\end{tabular}
\begin{tabular}{|c|c|}
\hline
Name & Bound~(GeV) \\ \hline \hline
D10 & 1410 \\
(Majorana) & 1677 \\ \hline
D11 & 538 \\ 
(Majorana) & 479 \\ \hline
D12 & 543 \\
(Majorana) & 484 \\ \hline
D13 & 678 \\ 
(Majorana) & 604 \\ \hline 
D14 & 681 \\ 
(Majorana) & 607 \\ \hline
\end{tabular}
\begin{tabular}{|c|c|}
\hline
Name & Bound~(GeV) \\ \hline \hline
C1-C2 & 8 \\ \hline
C3 & 575 \\ \hline
C4 & 556 \\ \hline
C5 & 402 \\ \hline
C6 & 572 \\ \hline \hline
R1-R2 & 7 \\ \hline
R3 & 338 \\ \hline
R4 & 680 \\ \hline
\end{tabular}

\caption{``Mono-everything'' collider bounds on the operators of Table~\ref{tab:listofoperators}, translated to the normalization used in this paper, as extracted for 10~GeV dark matter from Ref.~\cite{Zhou:2013fla}. Bounds on Majorana fermion and real scalar dark matter are extrapolated from listed bounds on Dirac fermion and complex scalar operators. \label{tab:collider}}
\end{table}

The direct collider bounds on operators that couple to quark mass (D1-D4, C1, C2, R1 and R2) are extremely weak. This is because the effective coupling is suppressed by the small mass of the proton's up, down, and strange quarks relative to LHC energies. However, if the dark matter also couples to the top quark proportional to its large mass, then there are additional signatures that can be searched for.

First, a large coupling at tree-level to top quarks generates a loop-level operator coupling the dark matter directly to gluons \cite{Haisch:2012kf}, similar to the effective coupling of the Higgs to gluons generated by the large Higgs-top coupling. That is, an operator like D1: $m_f \Lambda^{-3}[\tilde{\chi}\chi] [\bar{f}f]$, will generate an effective operator like D11: $\alpha_S (\Lambda')^{-3}[\tilde{\chi}\chi] [G_{\mu\nu}G^{\mu\nu}]$. In the limit of infinite top mass, the scale $\Lambda'$ is simply a loop-factor down from the effective-operator scale $\Lambda$. However, the $p_T$ cut required by the mono-jet searches make the infinite top-mass limit inappropriate and reduces the scale $\Lambda$ that can be ruled out from collider bounds. In Figs.~\ref{fig:signal}, \ref{fig:signal2}, \ref{fig:others}, and \ref{fig:others2}, I include the limits from the loop-induced operators from Ref.~\cite{Haisch:2012kf}, derived using $\sim 5$~fb$^{-1}$ of luminosity from the 7~TeV run at the LHC.

Secondly, large couplings to top (and bottom) quarks allow for searches in the $t\bar{t}+\slashed{E}_T$ and $b$-rich channels \cite{Lin:2013sca}. These searches do not require that the top-coupling live in the effective operator regime. However, using the published data, this approach provides somewhat weaker bounds than the top-loop gluon-induced operators, and so I do not include them in the exclusion plots.  

Note that the assumption that the effective operator formalism still applies translates into the assumption that the dark matter interaction is not mediated by a light particle. If a light mediator exists, as has been suggested as a possible particle physics model of the CoGeNT signal \cite{Hooper:2012cw}, then the mono-everything bounds are significantly weaker than reported. 

In addition, the validity of the effective operator formalism begins to break down at low scales $\Lambda$. Integrating out the heavy physics requires that the momentum transfer $Q_{\rm tr}$ in the collider pair-production be less than the scale $\Lambda$. The relevant momentum transfer depends on the exact experimental cuts, but for light dark matter, the regime of validity generically extends down to $Q_{\rm tr} \gtrsim 240~\mbox{GeV}$ \cite{Busoni:2013lha}, using the least-constraining jet $p_T > 110$~GeV selection requirement of Ref.~\cite{Zhou:2013fla}. Converting this to a domain of validity on $\Lambda$ requires knowledge of the couplings between the dark matter/Standard Model particles and the heavy mediator and is model-dependent.  Alternatively, the domain of validity for the effective formalism can be estimated by requiring the unitarity of forward $q\bar{q}$ scattering \cite{Shoemaker:2011vi,Fox:2012ee}. For operators that do not couple proportional to quark masses, this estimate suggests that the formalism breaks down for $\Lambda$ on the order of 100~GeV. Applying this to scalar and pseudo-scalar operators suggests that the effective operator regime extends to much lower scales, but further work is required here.

\section{Direct Detection through effective operators \label{sec:dd}}

In this Section, I consider the operators from Table~\ref{tab:listofoperators} that can provide direct detection signals which are not suppressed by powers of $q^2$ or $v^2$. I restrict myself to the CDMS-Si and CoGeNT signal regions. The best fit to the DAMA/LIBRA annual modulation, assuming a standard dark matter halo model with a Maxwell-Boltzmann velocity distribution, requires a significantly larger cross section than the CoGeNT/CDMS-Si best-fit. While it is conceivable that deviations from the standard halo model (see {\it e.g.}~Refs.~\cite{DelNobile:2013cta,Frandsen:2013cna,Mao:2013nda} for discussion of these effects), isospin-violating couplings, or unresolved experimental uncertainties could bring all three experimental results into alignment, such possibilities are beyond the scope of this paper. For similar reasons, I ignore the XENON100 bounds, which are in tension with the positive CoGeNT and CDMS-Si results. 

The best fit for the CoGeNT region at 90\% confidence has a dark matter particle with mass and spin-independent elastic nucleon cross section of 
\begin{eqnarray}
(m_\chi)_{\rm CoGeNT} & = & 6.95-9.02~\mbox{GeV}, \\
(\sigma^{\rm SI})_{\rm CoGeNT} & = & (2.47-4.73)\times 10^{-41}~\mbox{cm}^2 = (6.34\times 10^{-14}-1.22\times 10^{-13})~\mbox{GeV}^{-2}, 
\end{eqnarray}
assuming isospin-conserving interactions with the nucleons. I take this best-fit region from Ref.~\cite{Kelso:2011gd}, which applies a correction to the CoGeNT results \cite{Aalseth:2011wp,Aalseth:2010vx} to account for additional surface event contamination. The CDMS-Si best fit region covers more of the mass/cross section parameter space, which is not surprising, as this region is based on only three events. The best fit region at 90\% confidence is \cite{Agnese:2013rvf}
\begin{eqnarray}
(m_\chi)_{\rm CDMS} & = & 5.64-20.76~\mbox{GeV}, \\
(\sigma^{\rm SI})_{\rm CDMS} & = & (9.50\times 10^{-43}-7.59\times 10^{-40})~\mbox{cm}^2 =(2.44\times 10^{-15}-1.95\times10^{-12})~\mbox{GeV}^{-2}.
\end{eqnarray}
CDMS-Si reports the maximum likelihood corresponds to a mass of 8.6 GeV and cross section of $1.9 \times 10^{-41}$~cm$^2$, which falls just outside the CoGeNT 99\% confidence region.

For each effective operator from Table~\ref{tab:listofoperators} that induces a direct detection cross section, I calculate the energy scale $\Lambda$ that would be necessary to provide the CoGeNT/CDMS-Si signals. For each operator I show the scale $\Lambda$ required for the observed signals compared to that required for a thermal cross section in the early Universe that gives the observed amount of dark matter, using the relic abundance formulae from Appendix \ref{app:thermal}. I also show the lower bounds on $\Lambda$ from indirect detection and collider bounds. For all but one set of operators, the $\Lambda$ required for direct detection results in too much dark matter from thermal freeze-out. Other than this one set of operators, there must be some other annihilation channel open to the dark matter. If that channel can be written in the form of an effective operator, it cannot be any that give direct detection, and so must be one of the remaining operators, which I consider in Section~\ref{sec:others}. I begin with spin-independent scattering, and then consider the spin-dependent case.

\subsection{Spin Independent Operators}

For a Lagrangian resulting in an isospin-conserving spin independent cross section, the nucleon cross section is
\begin{eqnarray}
\sigma_{\rm fermion} & = & \frac{x \mu^2}{\pi} f^2, \\
\sigma_{\rm scalar} & = & \frac{x\mu^2}{4 \pi m_\chi^2} f^2,
\end{eqnarray}
where $\mu$ is the reduced mass of the nucleon/dark matter system, $x = 1(4)$ for Dirac (Majorana) fermions or complex (real) scalars, and $f$ is the effective nucleon coupling. For scalar operators coupling to quarks, the effective couplings for protons and neutrons are
\begin{equation}
f_{p,n} = \sum_{q=u,d,s} \xi_q f^{p,n}_q \frac{m_{p,n}}{m_q}+\frac{2}{27} f_{TG}^{p,n} \sum_{q=c,b,t} \xi_q \frac{m_{p,n}}{m_q}, \label{eq:scalarf}
\end{equation}
where $\xi_q$ is the quark/dark matter coupling from the effective Lagrangian, $m_q$ is the quark mass, and $f_q^{p,n}$ and $f_{TG}^{p,n}$ are proportional to quark expectation operators $\langle \bar{q}q\rangle$. These values are extracted from QCD lattice simulations \cite{Young:2009zb,Giedt:2009mr,Toussaint:2009pz}, and I adopt the values of Ref.~\cite{Fitzpatrick:2010em} (see also Ref.~\cite{Belanger:2008sj}). For the level of accuracy relevant to this paper, there is no significant difference between the proton and neutron $f_{p,n}$ \cite{Young:2009zb,Toussaint:2009pz,Giedt:2009mr,Fitzpatrick:2010em}. For the operator D1, $\xi_q = m_q/\Lambda^3$; while the scalar dark matter operators C1 and R1 have $\xi_q = m_q/\Lambda^2$. 

For direct detection mediated by a vector operator coupled to quarks, such as D5, only the valence quarks contribute. Here, the coupling to protons and neutrons is 
\begin{equation}
f_{p,n} = \sum_{q=u,d} n_q \xi_q,
\end{equation}
with $n_q$ the number of valence quarks of flavor $q$, and $\xi_q$ the Lagrangian couplings. For operator D5, $\xi_q = \Lambda^{-2}$, and for both neutrons and protons  $f_{p,n} = 3\Lambda^{-2}$, due to the the flavor universal coupling.

For operators that couple to gluons (such as D11), the effective nucleon coupling is given by \cite{Shifman:1978zn,Fox:2011pm}
\begin{eqnarray}
f_g & = & \xi_g \langle N |\alpha_S G_{\mu\nu}G^{\mu\nu} |N\rangle \\
 & = & \xi_g\left[-\frac{8\pi}{9}m_N\left(1-\sum_{u,d,s} f_q^N \right)\right],
\end{eqnarray}
where $\xi_g$ is the prefactor of the operator modulo a factor of $\alpha_S$ (so, for example, $\xi_g = \Lambda^{-3}$ for operator D11), and the $f_q^N$ are the same as those used in the scalar couplings of Eq.~\eqref{eq:scalarf}.

\begin{figure}[ht]
\includegraphics[width=0.25\columnwidth]{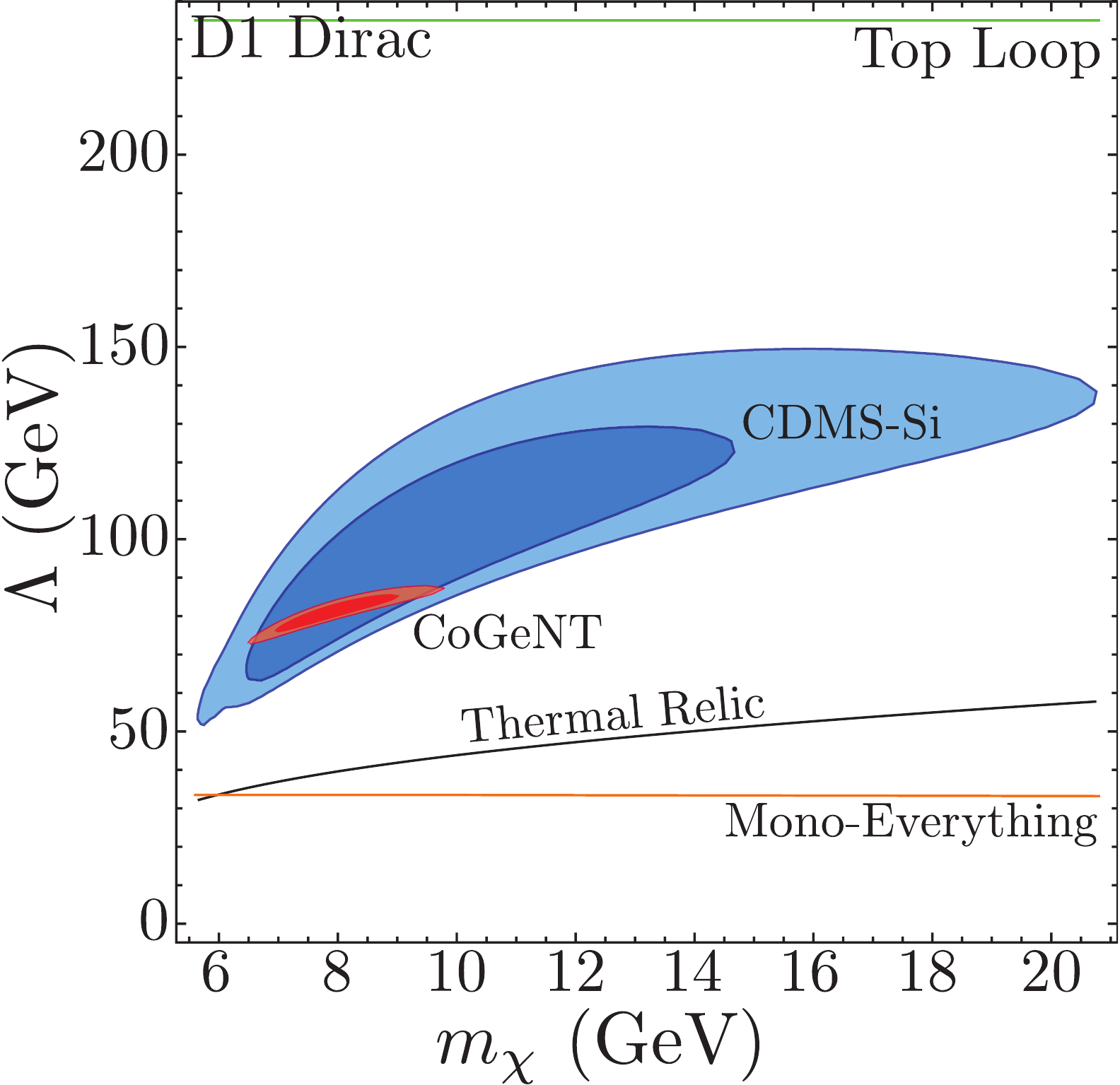}~\includegraphics[width=0.25\columnwidth]{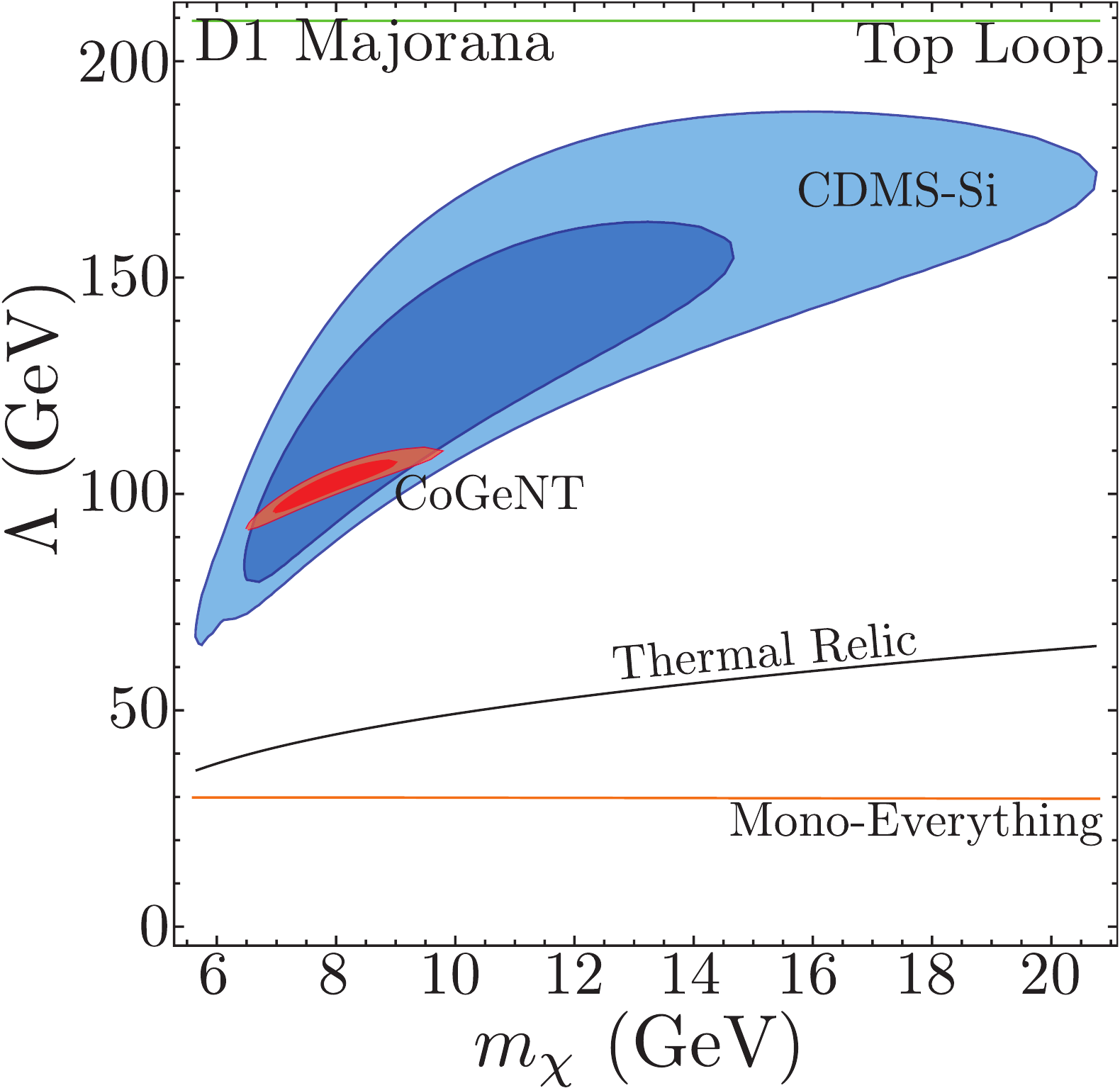}~\includegraphics[width=0.242\columnwidth]{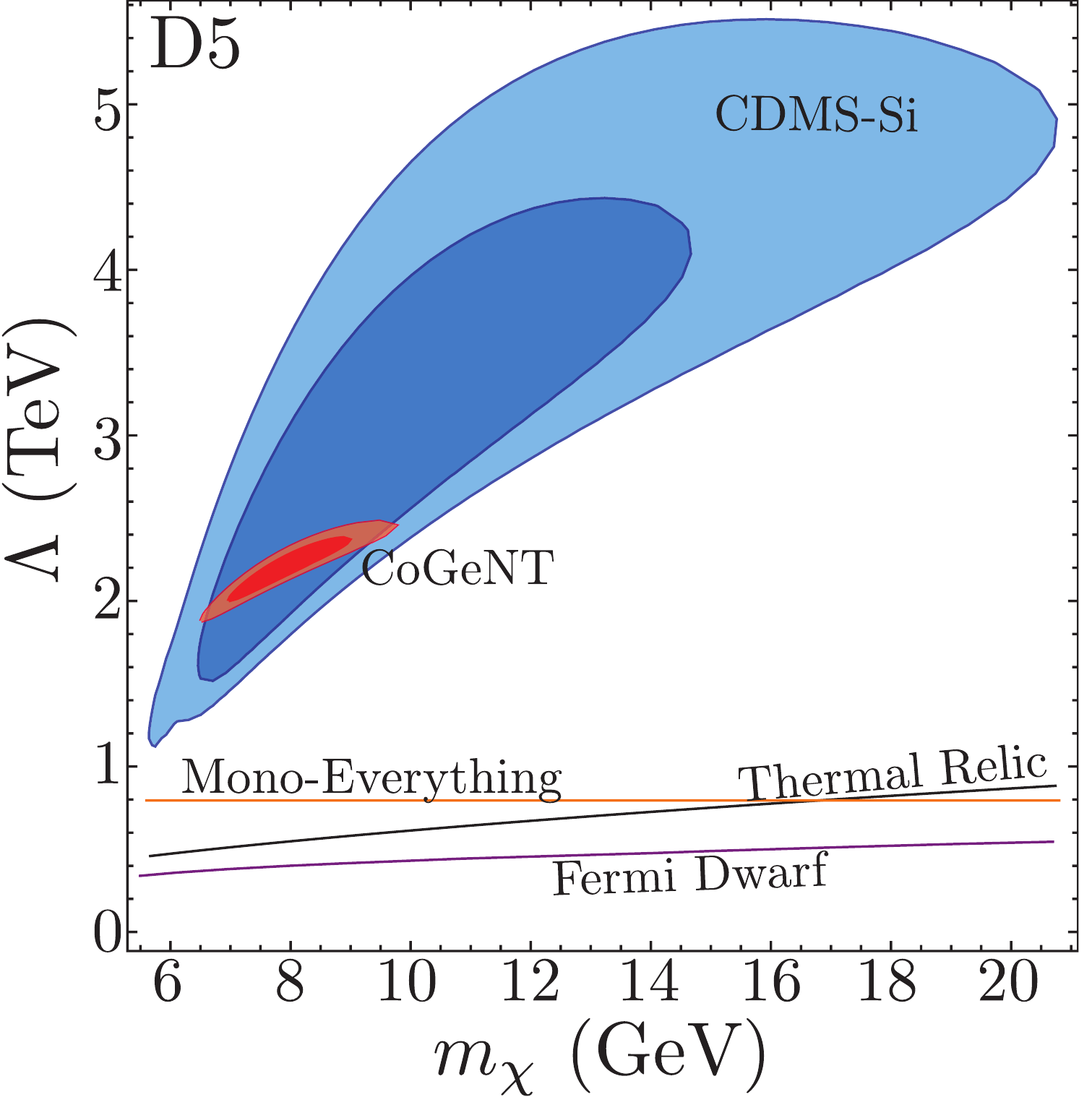}~\includegraphics[width=0.25\columnwidth]{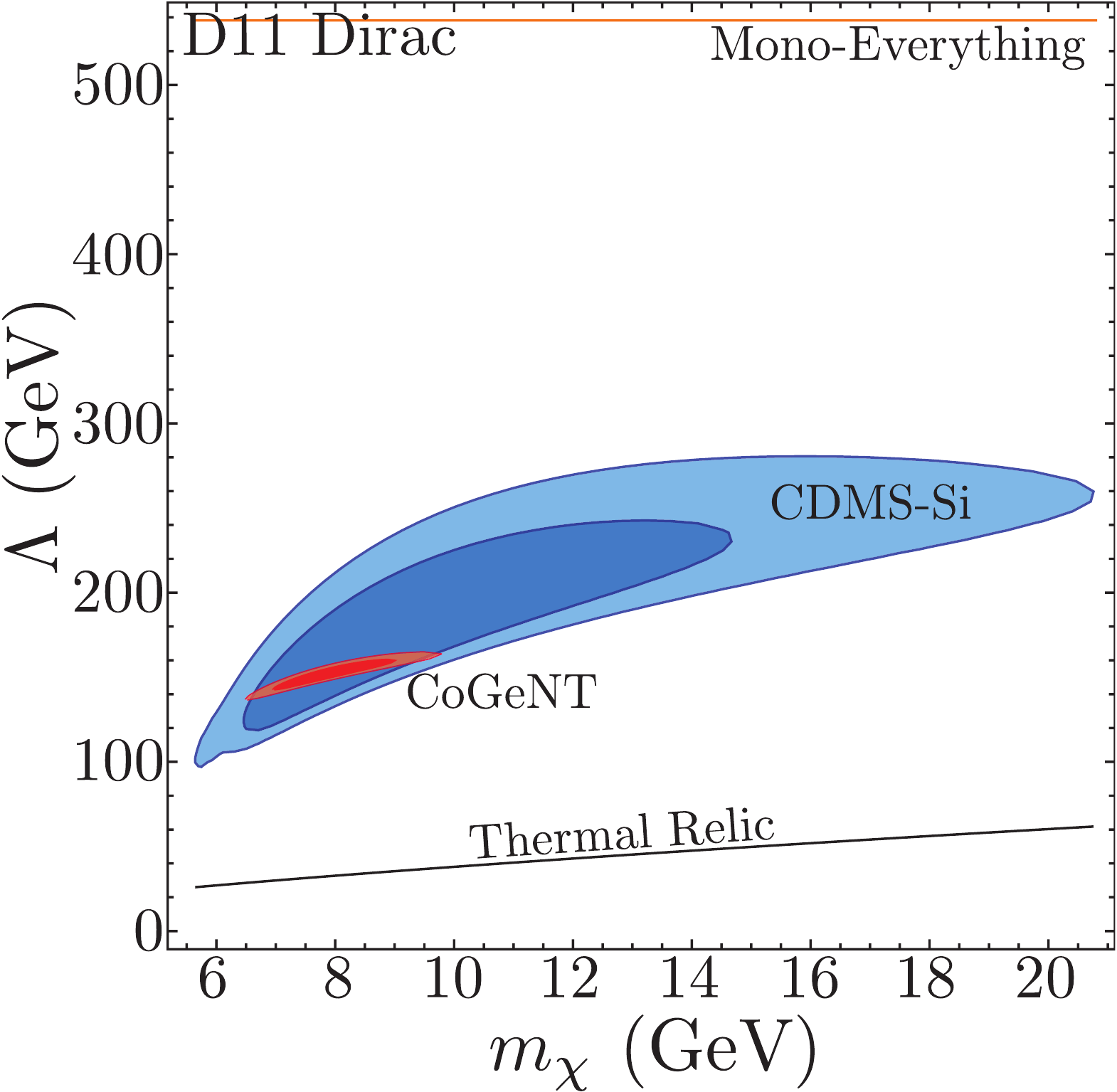}
\includegraphics[width=0.25\columnwidth]{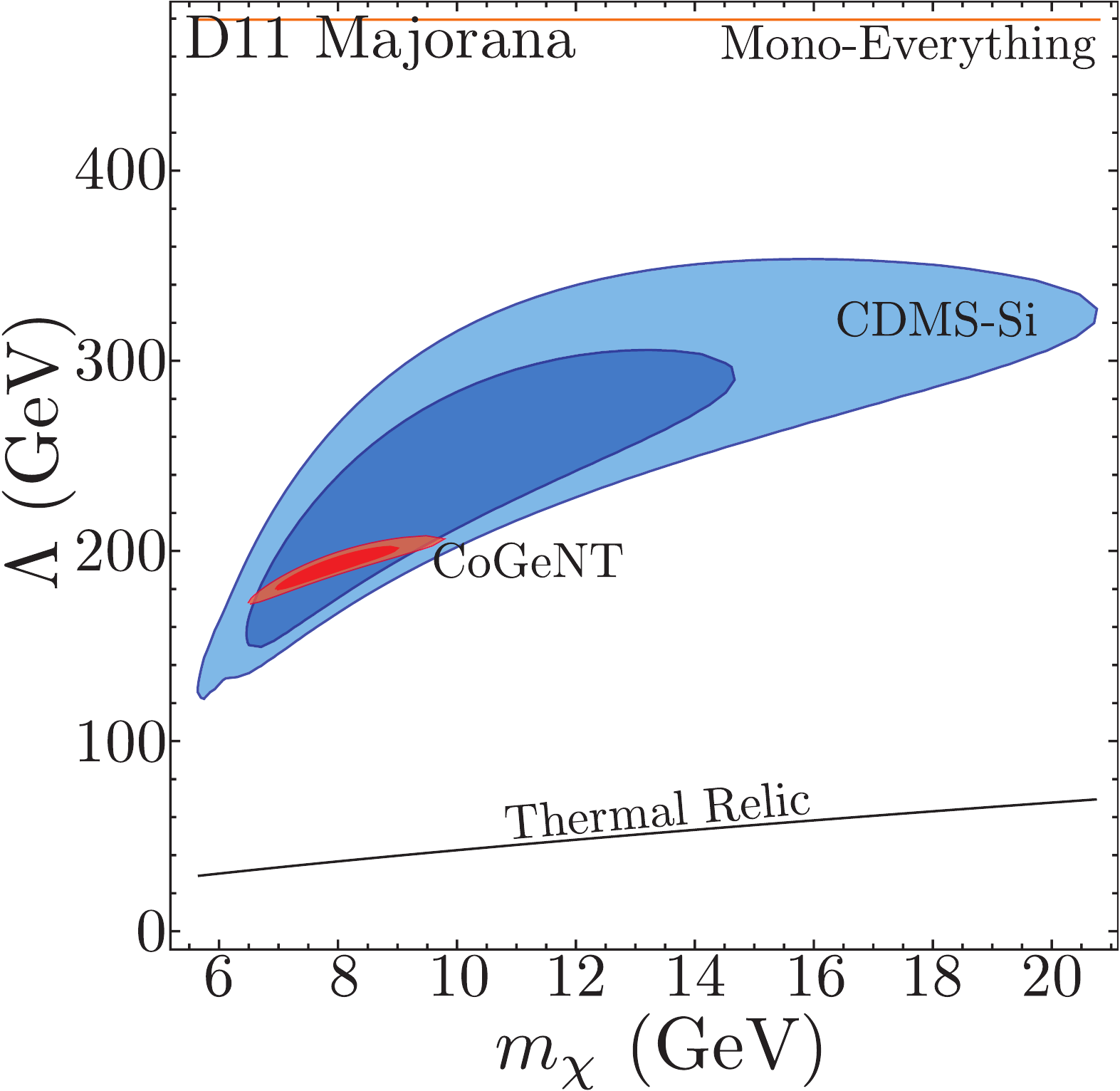}~\includegraphics[width=0.25\columnwidth]{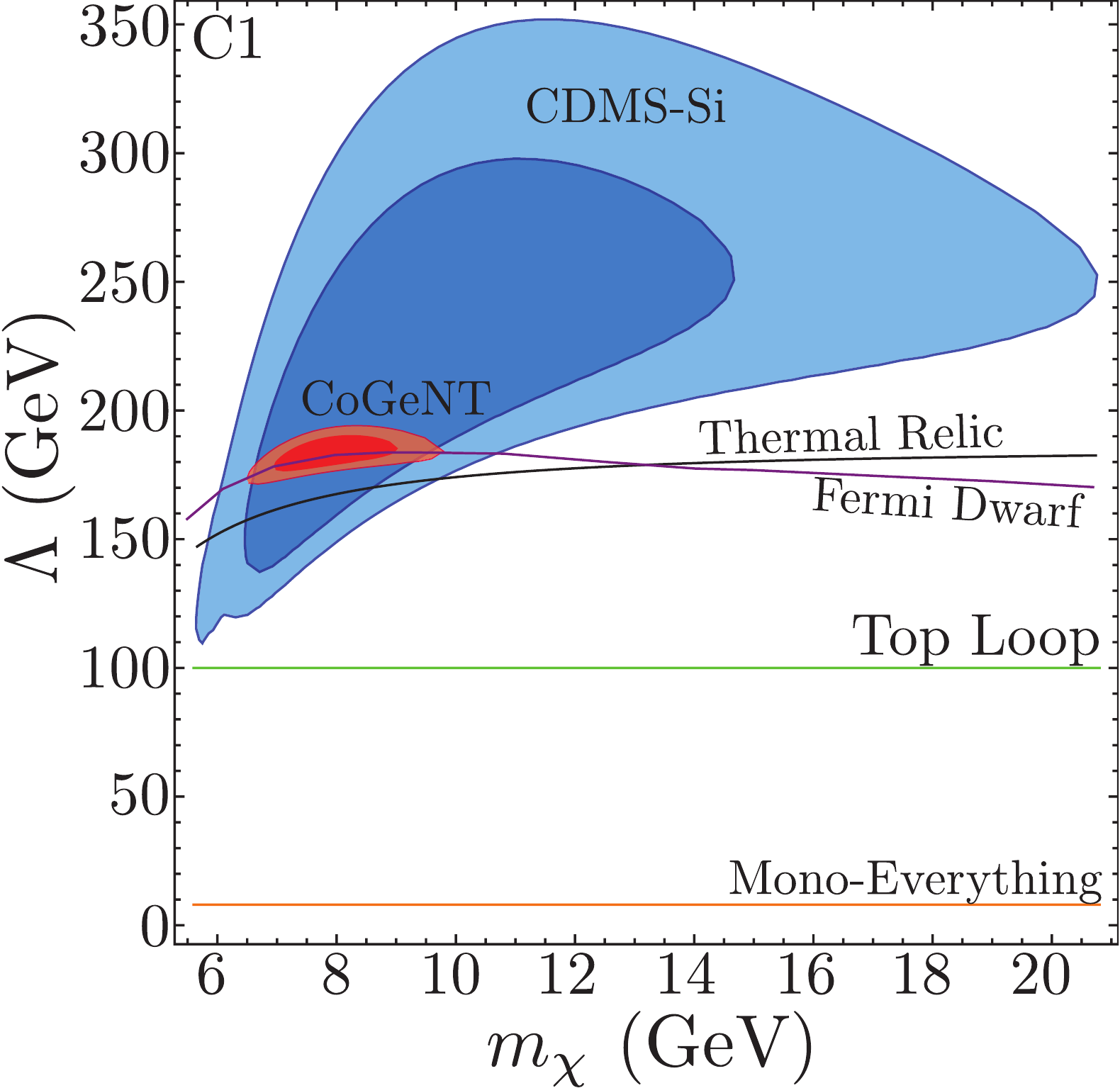}~\includegraphics[width=0.243\columnwidth]{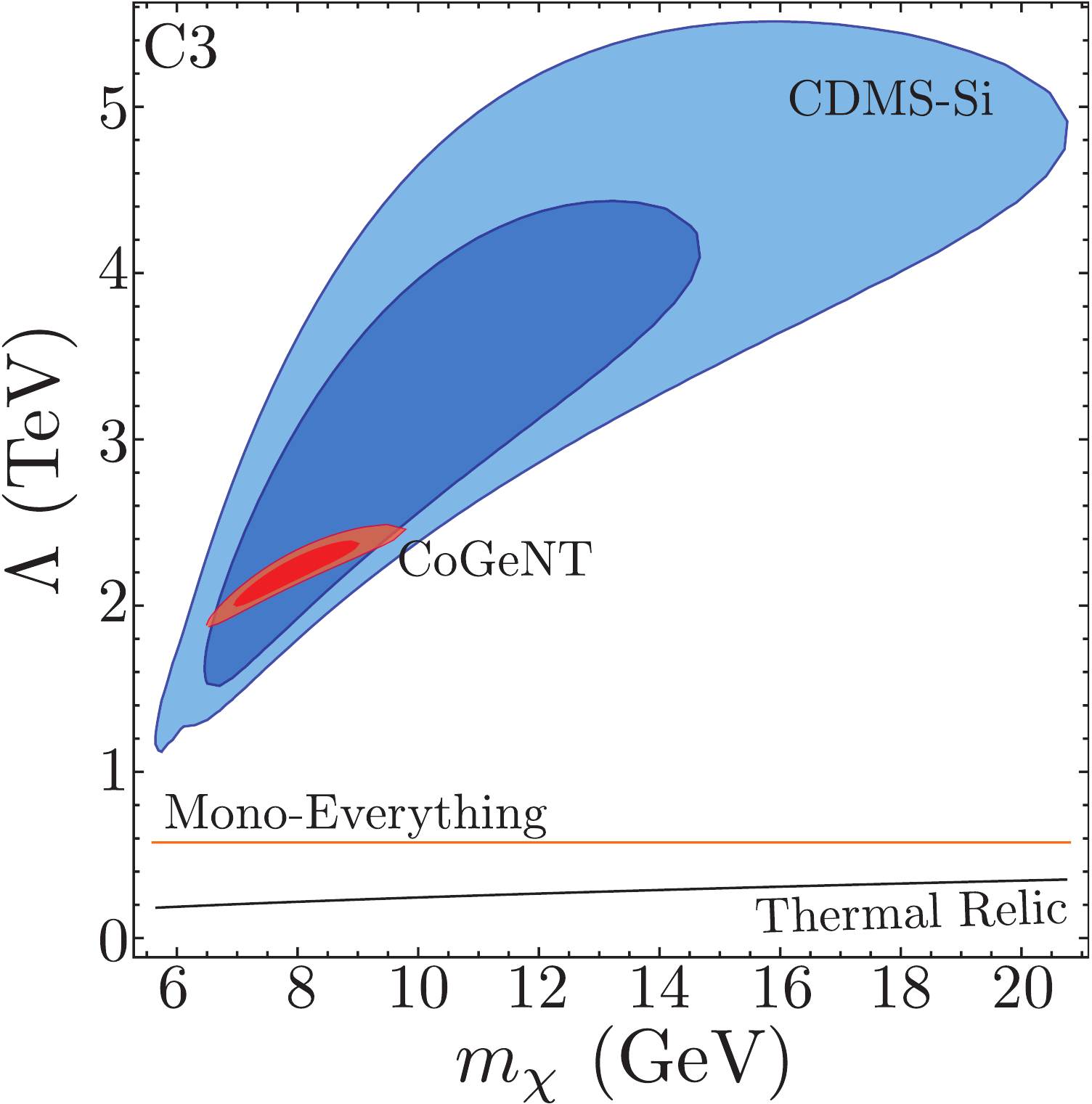}~\includegraphics[width=0.25\columnwidth]{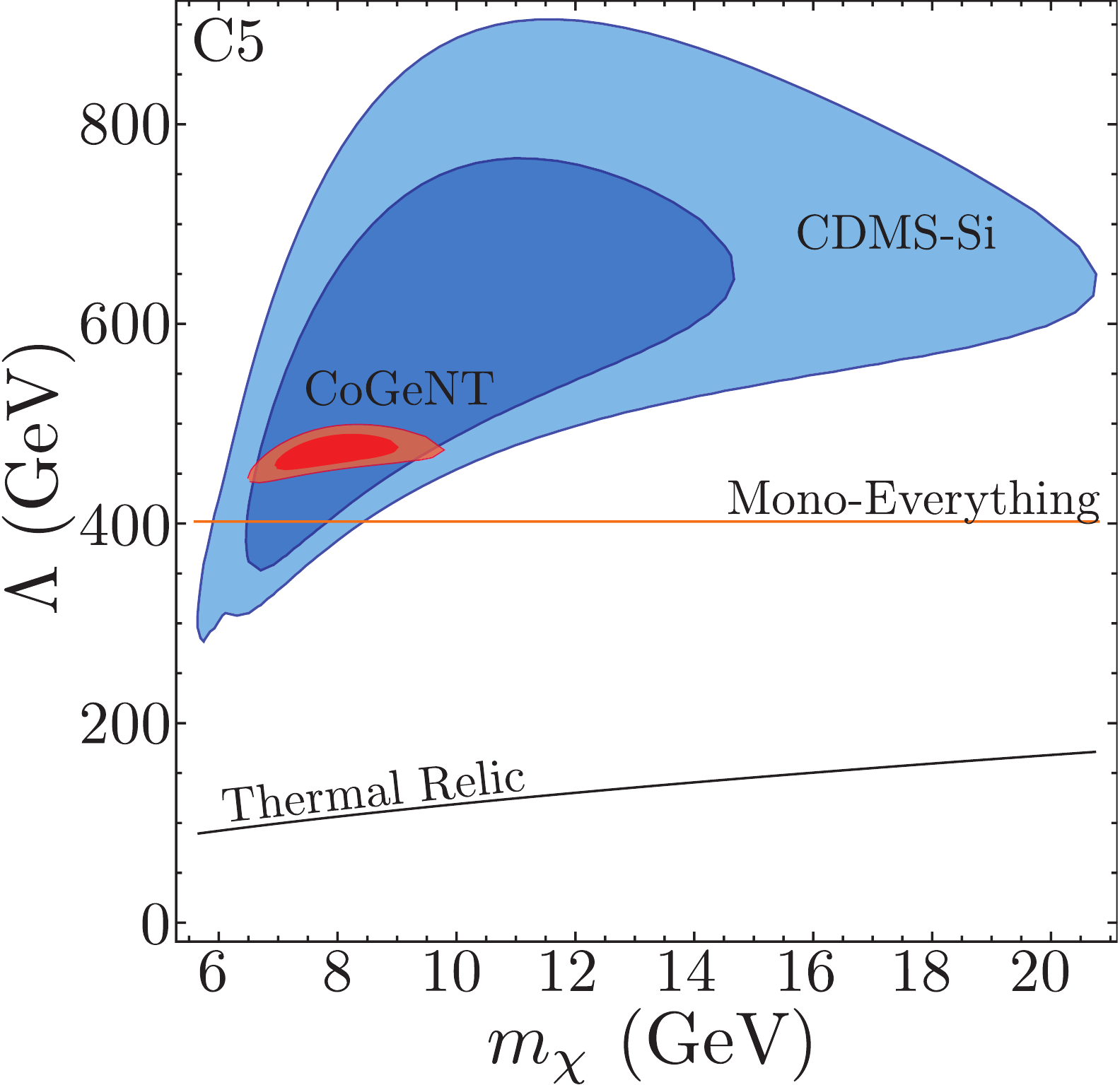}
\includegraphics[width=0.25\columnwidth]{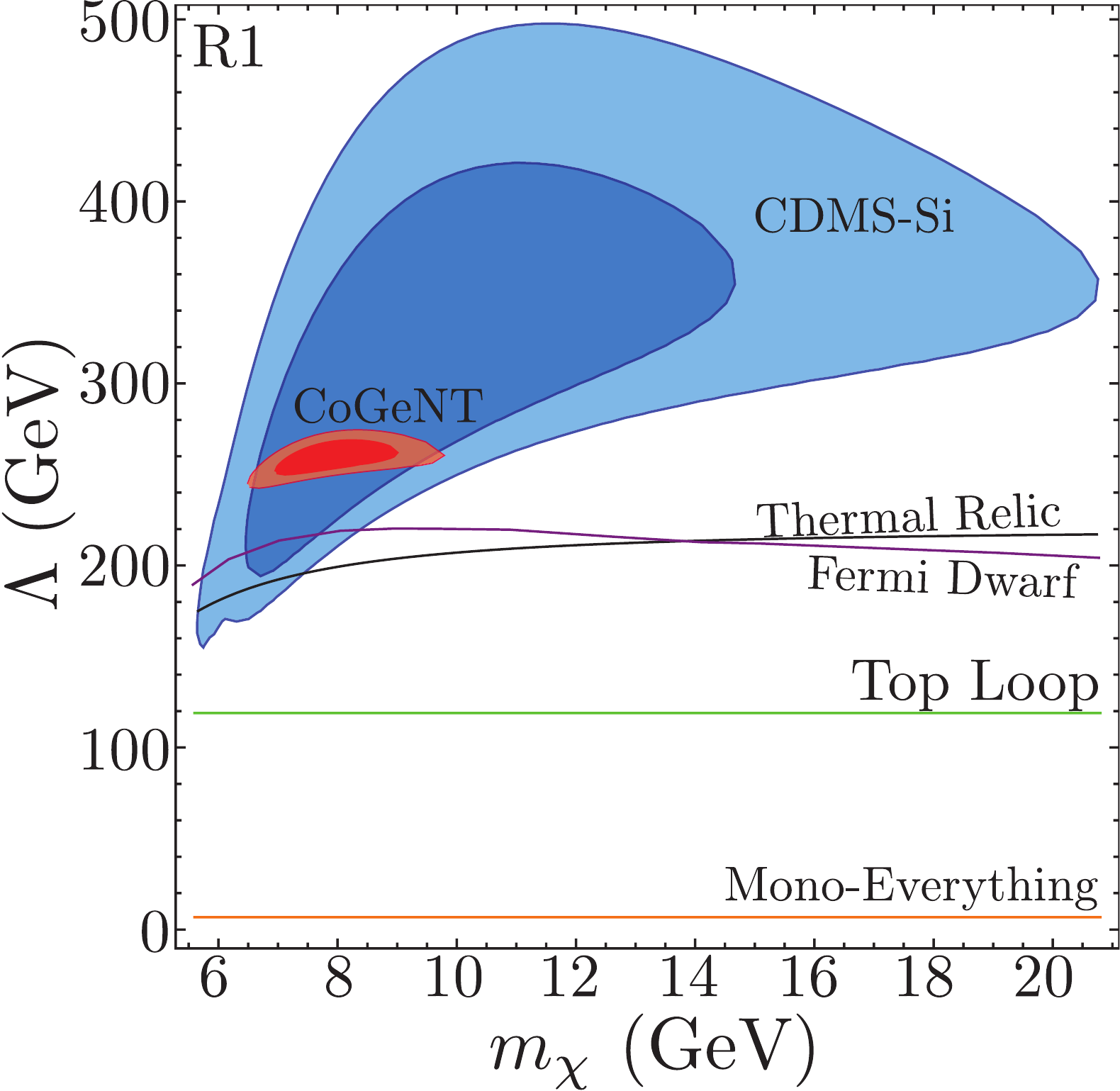}~\includegraphics[width=0.25\columnwidth]{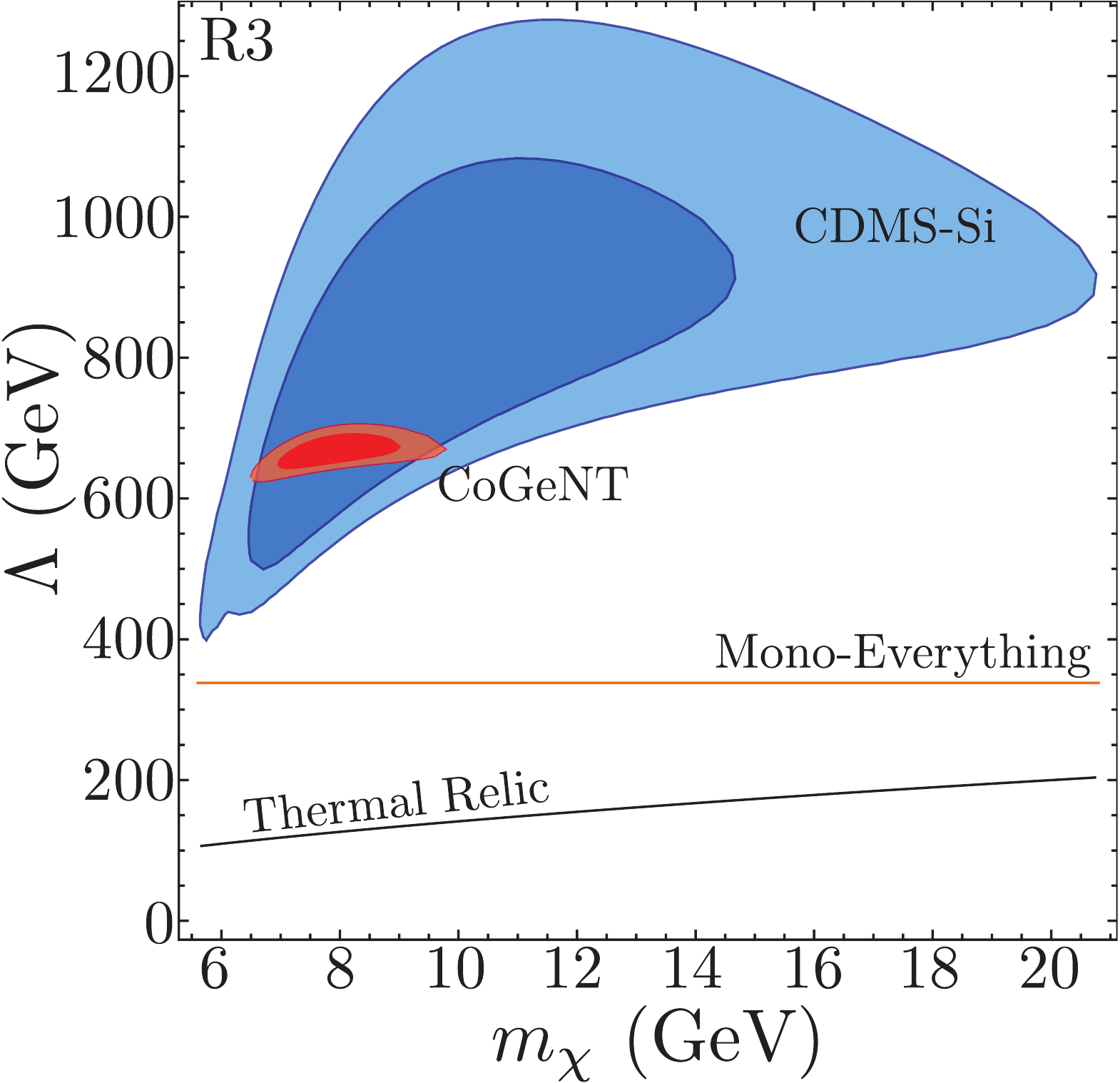}

\caption{Effective operator energy scale $\Lambda$ for operators from Table~\ref{tab:listofoperators} necessary to give a spin-independent cross section which explains the CoGeNT (red, 90\% and 99\% CL) and CDMS-Si (blue, 68\% and 90\% CL) signal regions as function of dark matter mass $m_\chi$. Also shown are the values of $\Lambda$ as a function of $m_\chi$ giving the correct thermal relic abundance of dark matter (black line), the lower limits on $\Lambda$ from the ``mono-everything'' collider searches \cite{Zhou:2013fla} (orange line), and the lower limit from the Fermi dwarf galaxy indirect detection searches \cite{Ackermann:2011wa} (purple line). For operators with couplings proportional to quark mass, the limits of Ref.~\cite{Haisch:2012kf}, derived from the top-loop induced production, are shown as a green line. \label{fig:signal}}
\end{figure}

In Fig.~\ref{fig:signal}, I plot the range of $\Lambda$ as a function of dark matter mass $m_\chi$ which provide the CoGeNT or CDMS-Si direct detection signal strength through spin-independent interactions. I plot also the $\Lambda$ values which would provide sufficient annihilation in the early Universe, using the thermal cross sections discussed in the previous section (and listed in Appendix~\ref{app:thermal}). I also display the lower bound on $\Lambda$ from the Fermi observations of dwarf galaxies \cite{Ackermann:2011wa,GeringerSameth:2011iw}, considering only annihilation into $b\bar{b}$ final states. Some operators have only velocity-suppressed annihilations or annihilations into only gluons, and so are not constrained by $b\bar{b}$ indirect detection results. Finally, I show the lower limit on $\Lambda$ for each operator from the ``mono-everything'' collider searches of Ref.~\cite{Zhou:2013fla}, as well as the bound derived from these searches assuming induced gluon-dark matter production for the effective operators with scalar couplings proportional to the Standard Model fermion mass, taken from Ref.~\cite{Haisch:2012kf}. I do not plot the weaker $b$- and $t$-enriched bounds from Ref.~\cite{Lin:2013sca}.

As can be seen, only operators C1 and R1 have regions broadly consistent with the CoGeNT and CDMS-Si results that also have a large enough annihilation cross section in the early Universe to produce the correct relic density through a thermal process, though this thermal value is in some tension with the null results of the Fermi dwarf search. Operator D11 is directly ruled out by the collider constraints, assuming the validity of the effective theory. For the other operators, the values of $\Lambda$ that would explain the direct detection results are too large compared to the relic abundance requirement. As a result, the thermal cross section these operators would induce is too small, and so dark matter would be overproduced. For these operators to explain the CoGeNT/CDMS-Si results, some other annihilation process would have to be involved. I will discuss the possible contributions from other effective operators in Section~\ref{sec:others}.

Returning to operators C1 and R1, which appear to allow for a thermal dark matter particle (in particular, either a complex or real scalar) giving the measured direct detection signal, one can see that the combined collider bounds are too weak to directly constrain the required values of $\Lambda$. The loop-induced constraints approach, but do not exclude these regions, while some of the regions (along with some of the thermal relic value) are excluded by the Fermi dwarf searches. This puts these operators in an interesting position: they are capable of providing both a thermal relic and the observed direct detection signal and would not yet have been seen in colliders or in annihilation in the Universe today. As they couple proportionally to the quark mass, one might suspect a relation to the Higgs mechanism; however, the upper limits on the invisible Higgs width as measured by the LHC \cite{Djouadi:2011aa} strongly constrain Higgs coupling to dark matter with mass less than half the Higgs mass. Removing the coupling to $b$-quarks would leave the CoGeNT and CDMS regions essentially unchanged while eliminating much of the Fermi dwarf constraint; however, this set of couplings would then be non-minimally flavor violating.

\subsection{Spin Dependent Operators}

For spin-dependent interactions the cross section for a target nuclei with spin $J$ and nucleon spin expectation values $\langle S_p \rangle$ and $\langle S_n\rangle$ is \cite{Jungman:1995df}
\begin{equation}
\sigma_{\rm SD} = \frac{x\mu^2}{\pi}\left[a_p\langle S_p\rangle+a_n\langle S_n\rangle \right]^2\frac{J+1}{J},
\end{equation}
where $x = 1(4)$ for Dirac (Majorana) fermions. For axial-vector operators of the form $[\bar{\chi} \gamma^\mu\gamma_5 \chi][\bar{f} \gamma_\mu\gamma_5 f]$, the nucleon couplings $a_p$ and $a_n$ are given in terms of the Lagrangian-level couplings $\xi_q$ and the nucleon spins $\Delta q$ \cite{Jungman:1995df} as
\begin{equation}
a_{p,n} = \sum_{u,d,s} \xi_q \Delta q^{p,n}.
\end{equation}
For operator D8, the $\xi_q$ are flavor universal, and equal to $\Lambda^{-2}$. The low-velocity expansions of tensor operators $[\bar{\chi} \sigma^{\mu\nu} \chi][\bar{f} \sigma_{\mu\nu} f]$ contains an axial-vector coupling, with an overall numerical factor of 2 \cite{Agrawal:2010fh}. Thus, the flavor universal $\xi_q$ are $2\Lambda^{-2}$ for D9.

\begin{figure}[ht]
\includegraphics[width=0.3\columnwidth]{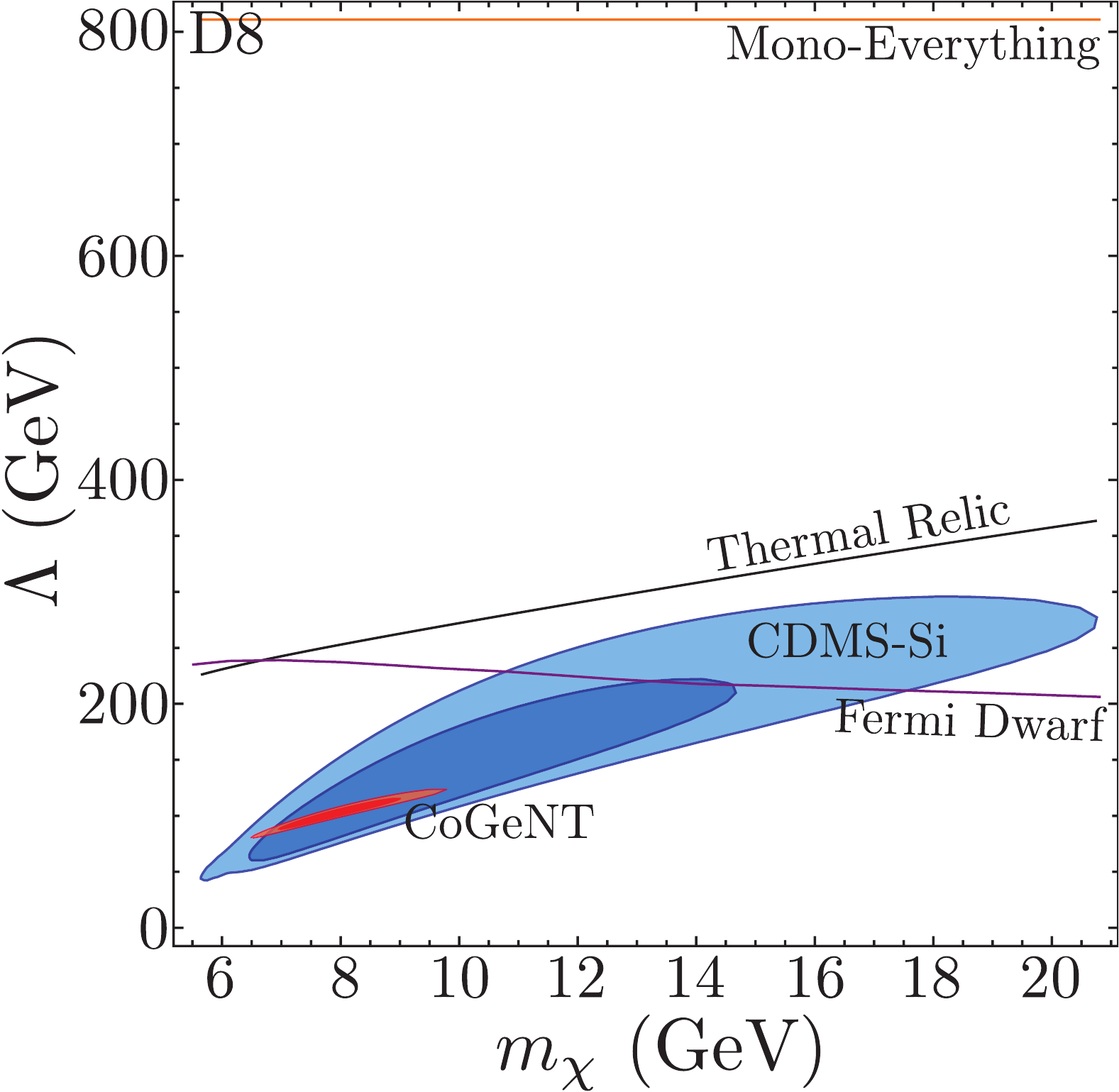}~\includegraphics[width=0.3\columnwidth]{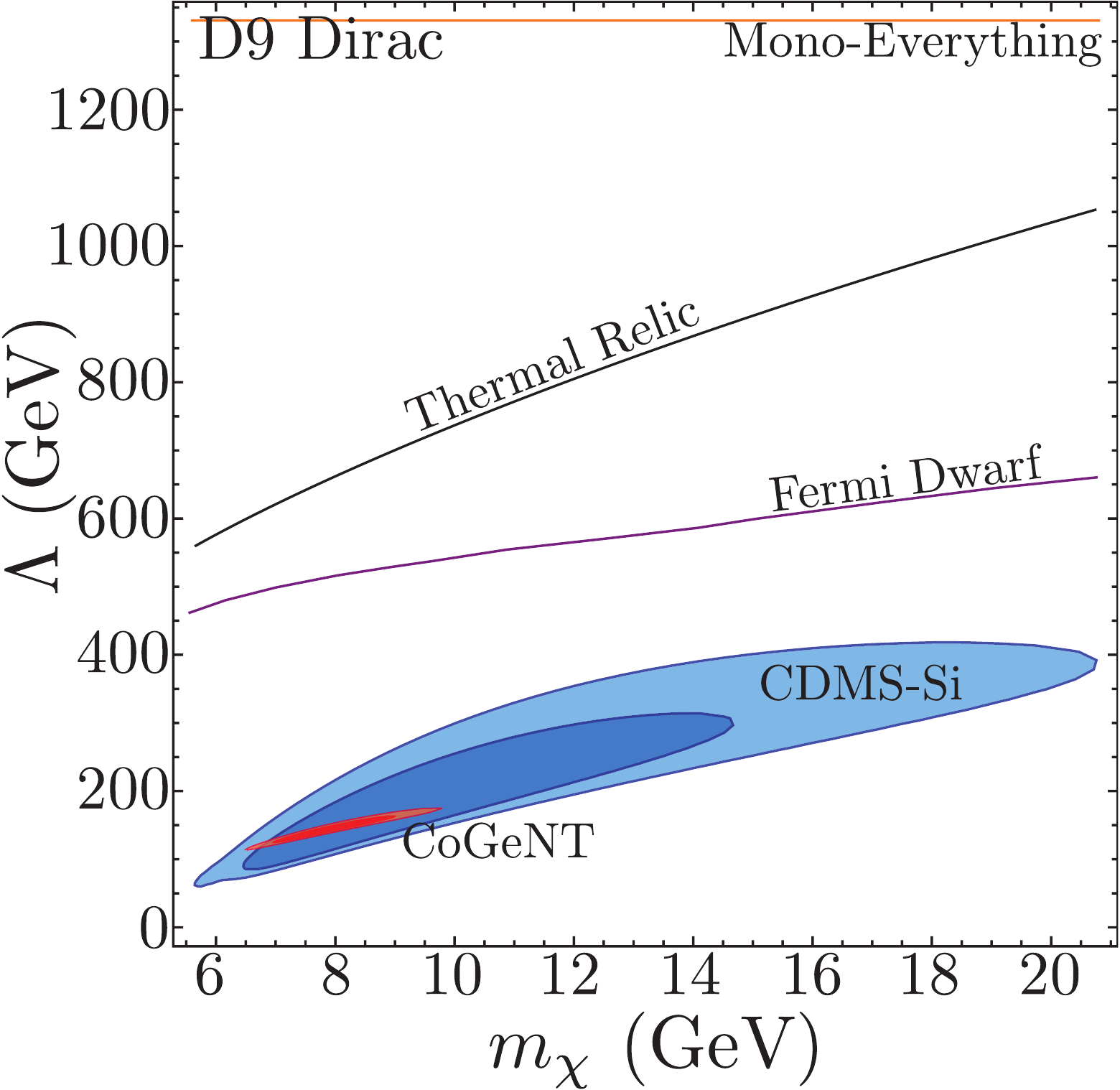}~\includegraphics[width=0.3\columnwidth]{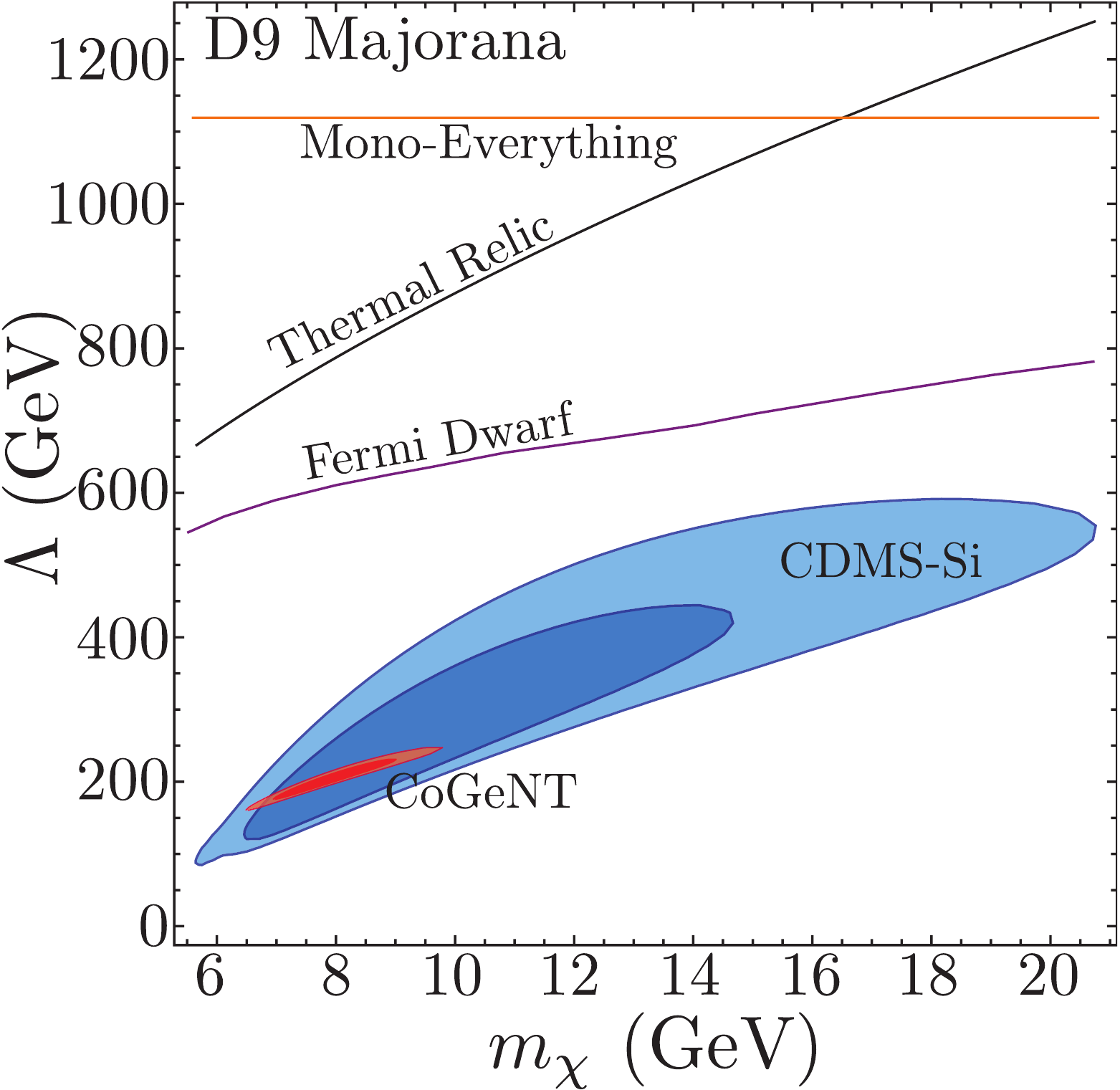}

\caption{Effective operator energy scale $\Lambda$ for operators from Table~\ref{tab:listofoperators} necessary to give a spin-dependent cross section which explains the CoGeNT (red, 90\% and 99\% CL) and CDMS-Si (blue, 68\% and 90\% CL) signal regions as function of dark matter mass $m_\chi$. Also shown are the values of $\Lambda$ as a function of $m_\chi$ giving the correct thermal relic abundance of dark matter (black line), the lower limit on $\Lambda$ from the ``mono-everything'' collider searches \cite{Zhou:2013fla} (orange line) and the lower limit from the Fermi dwarf galaxy indirect detection searches \cite{Ackermann:2011wa} (purple line). \label{fig:signal2}}
\end{figure}

Due to the flavor-universality assumption, the spin-dependent couplings will be equal for both proton and neutrons. This will bring the CoGeNT and CDMS-Si signals in conflict \cite{Buckley:2013gjo} with the PICASSO \cite{Archambault:2012} and COUPP \cite{Behnke:2012} null-results. However, even ignoring these constraints, the effective operator formalism does not yield a consistent picture for a spin-dependent interpretation of the low-mass dark matter results. As shown in Fig.~\ref{fig:signal2}, not only do none of the operators that result in spin-dependent cross sections have values of $\Lambda$ which produce correct direct detection and relic abundance cross sections, but the values of $\Lambda$ required for the CoGeNT/CDMS-Si and early Universe relics are already directly ruled out by the collider and indirect detection constraints. Therefore, if the CoGeNT and CDMS-Si results are due to spin-dependent interactions, it cannot be written as an effective operator.

\section{Additional Operators \label{sec:others}}

In the previous Section, I showed that only operators C1 and R1 can explain both the direct detection results from CoGeNT and CDMS-Si while providing the correct abundance of dark matter from thermal production. Of the remaining operators however, only one (operator D11) has direct detection regions unambiguously excluded by collider constraints (though D1 is constrained by the loop-induced collider signals). Therefore, it is possible that the signals reported by the direct detection experiments proceed through one operator, which provides only a subdominant annihilation cross section in the early Universe, while the remaining required annihilation occurs through additional effective operators that do not result in direct detection signals unsuppressed by powers of the dark matter velocity or momentum transfer. Additional operators could also reduce the tension with the indirect detection results for operators C1 and R1, by decoupling the relationship between the scale $\Lambda$ required for a thermal relic with that required for the direct detection signal.

As a specific example, consider Dirac dark matter inducing a spin-independent direct detection cross section through operator D5. To produce the observed signal, the operator must be suppressed by $\Lambda \sim 2$~TeV (taking the center of the CoGeNT best-fit region). This large of a value for $\Lambda$ cannot be excluded by collider bounds, though we see that this scale is not consistent with a thermal relic produced in the early Universe. However, one could argue that another operator coupling Dirac dark matter to Standard Model fermions could produce the required early Universe annihilation. This operator cannot produce additional direct detection signals, so it must be some combination of D2-D4, D6, D7, D10, and/or D12-14. Each of these operators has a particular value of $\Lambda$ that will give the correct thermal relic abundance (adding several operators together produces an ${\cal O}(1)$ change in the required cross section, which considering that $\sigma \propto \Lambda^{-4}$ or $\Lambda^{-6}$, has a negligible impact on these conclusions). Each of these operators also has an upper limit on $\Lambda$ from collider searches. It turns out that, in each case for the Dirac dark matter, the collider searches exclude the thermal value. Thus, even allowing additional annihilation through other effective operators does not save direct detection of thermal relics through operator D5. Similar arguments apply for the other direct detection operators, again with the notable exceptions of C1 and R1.

\begin{figure}[ht]
\includegraphics[width=0.25\columnwidth]{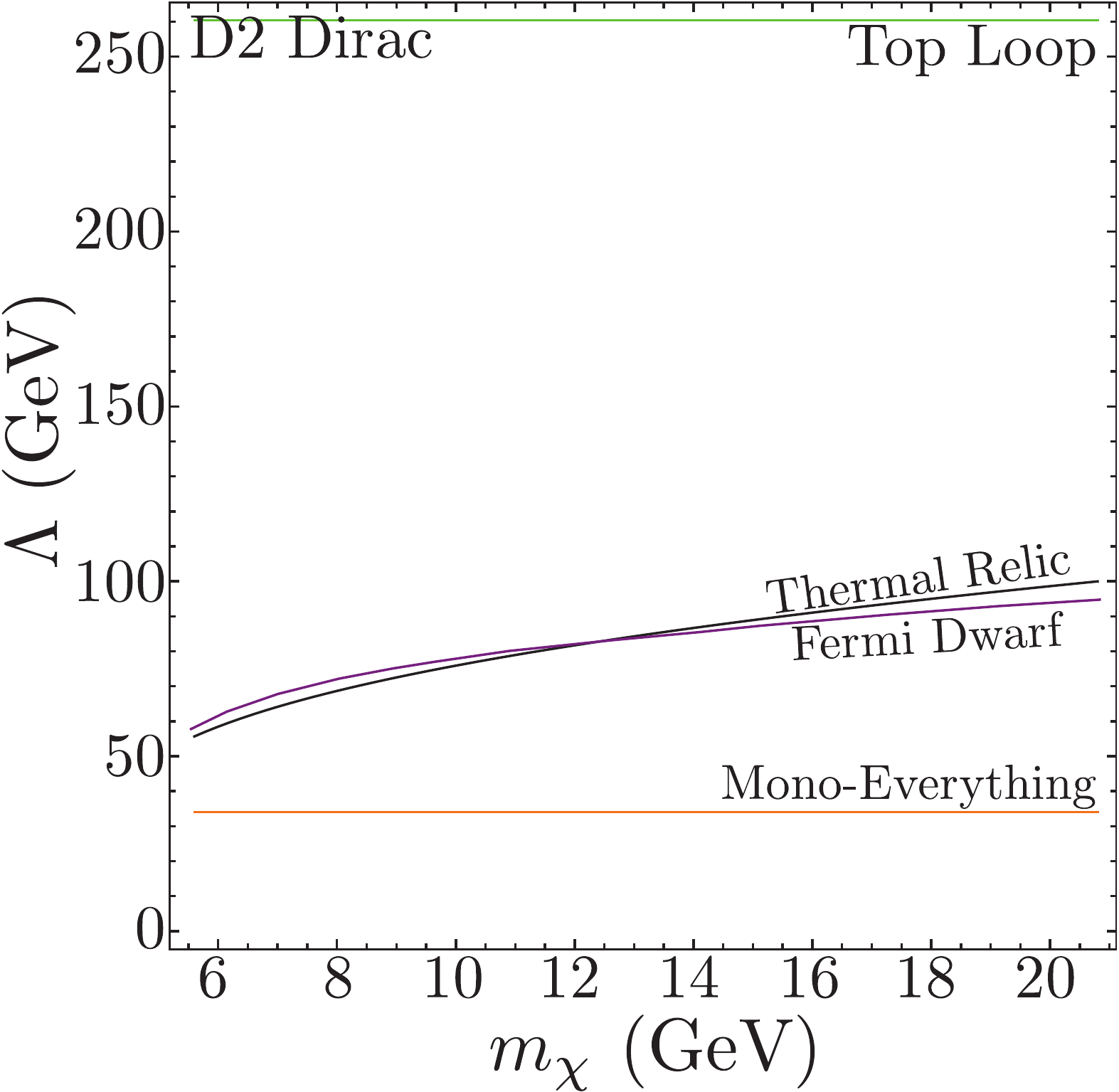}~\includegraphics[width=0.25\columnwidth]{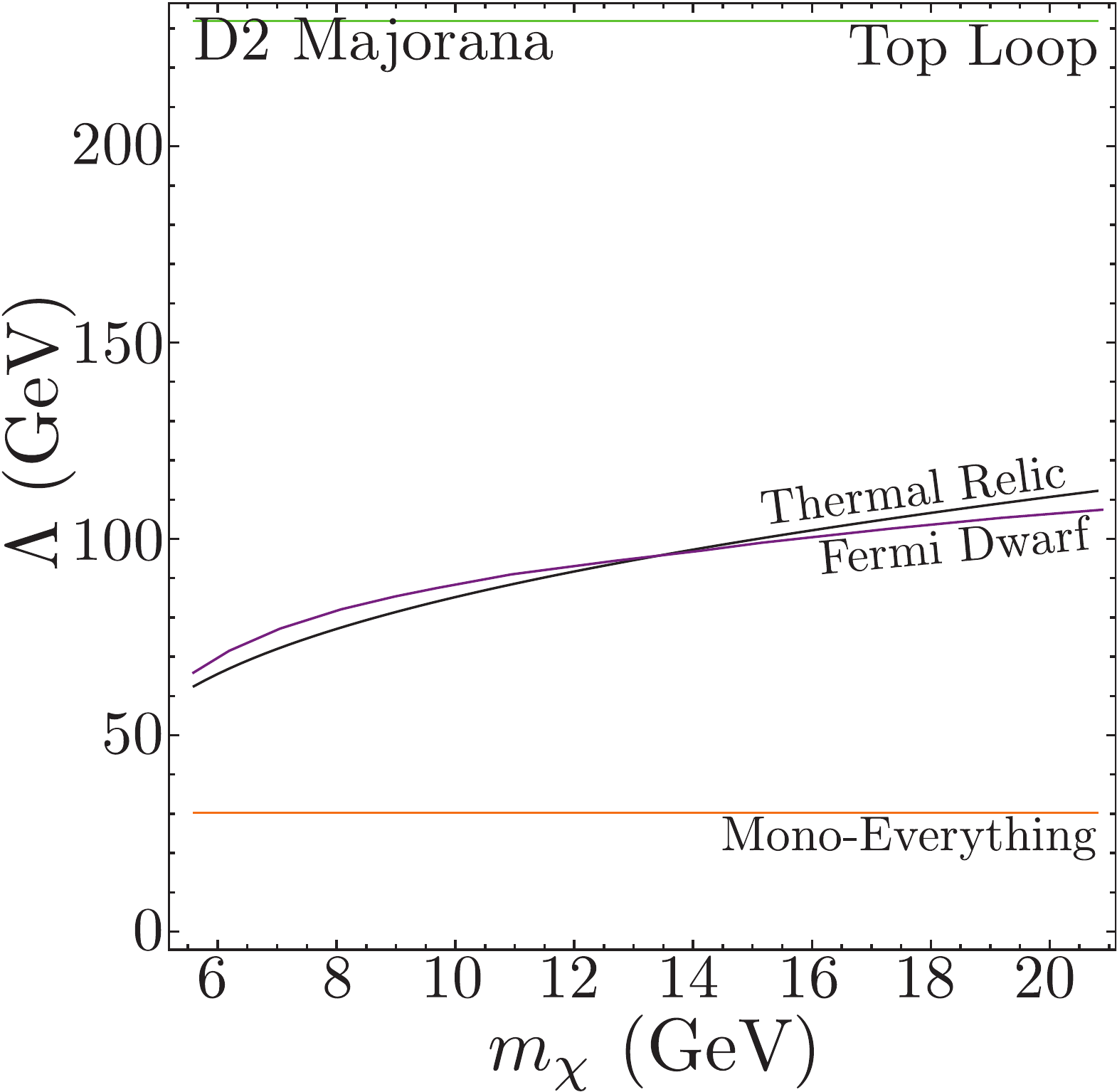}~\includegraphics[width=0.25\columnwidth]{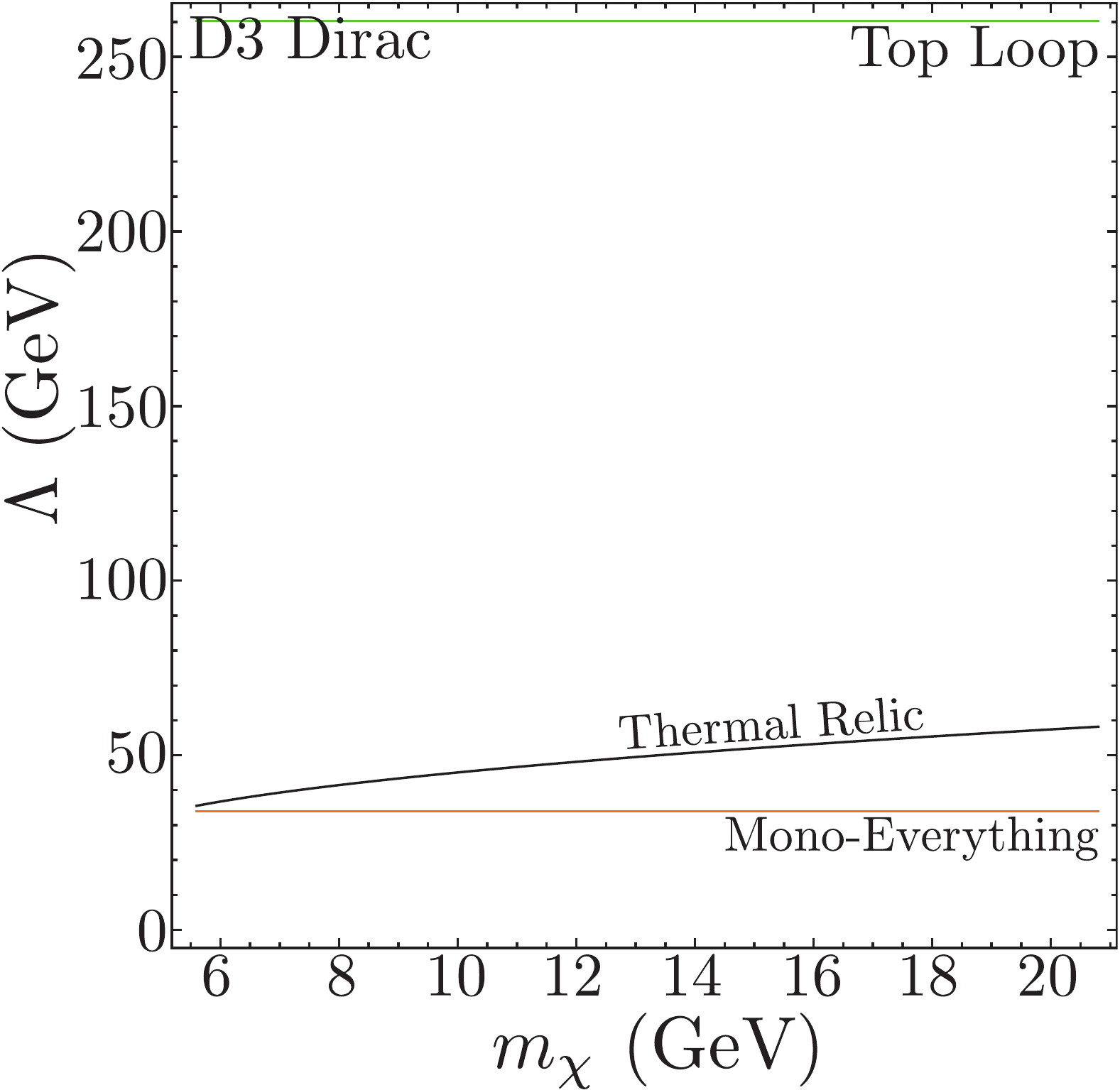}~\includegraphics[width=0.25\columnwidth]{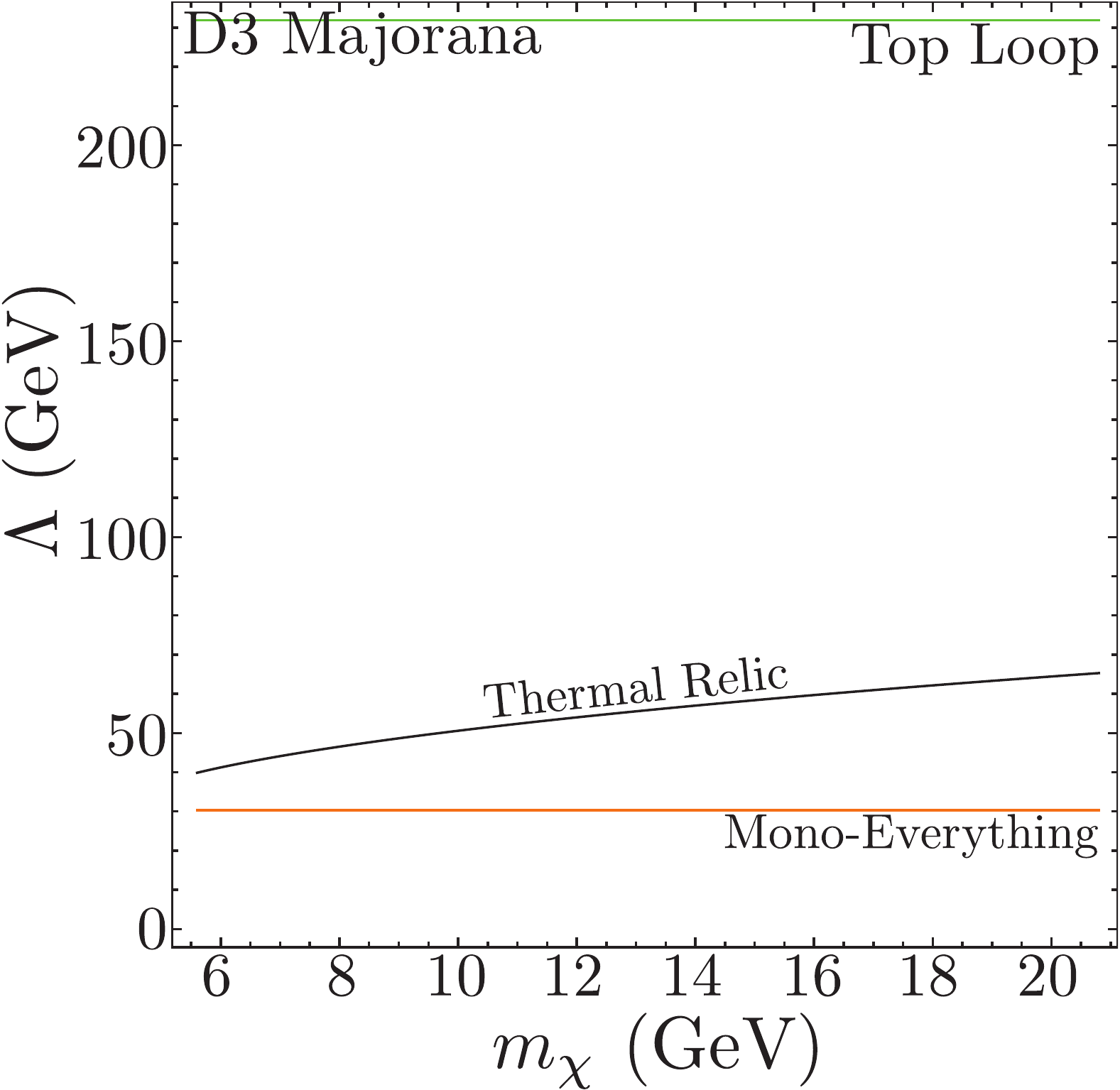}
\includegraphics[width=0.25\columnwidth]{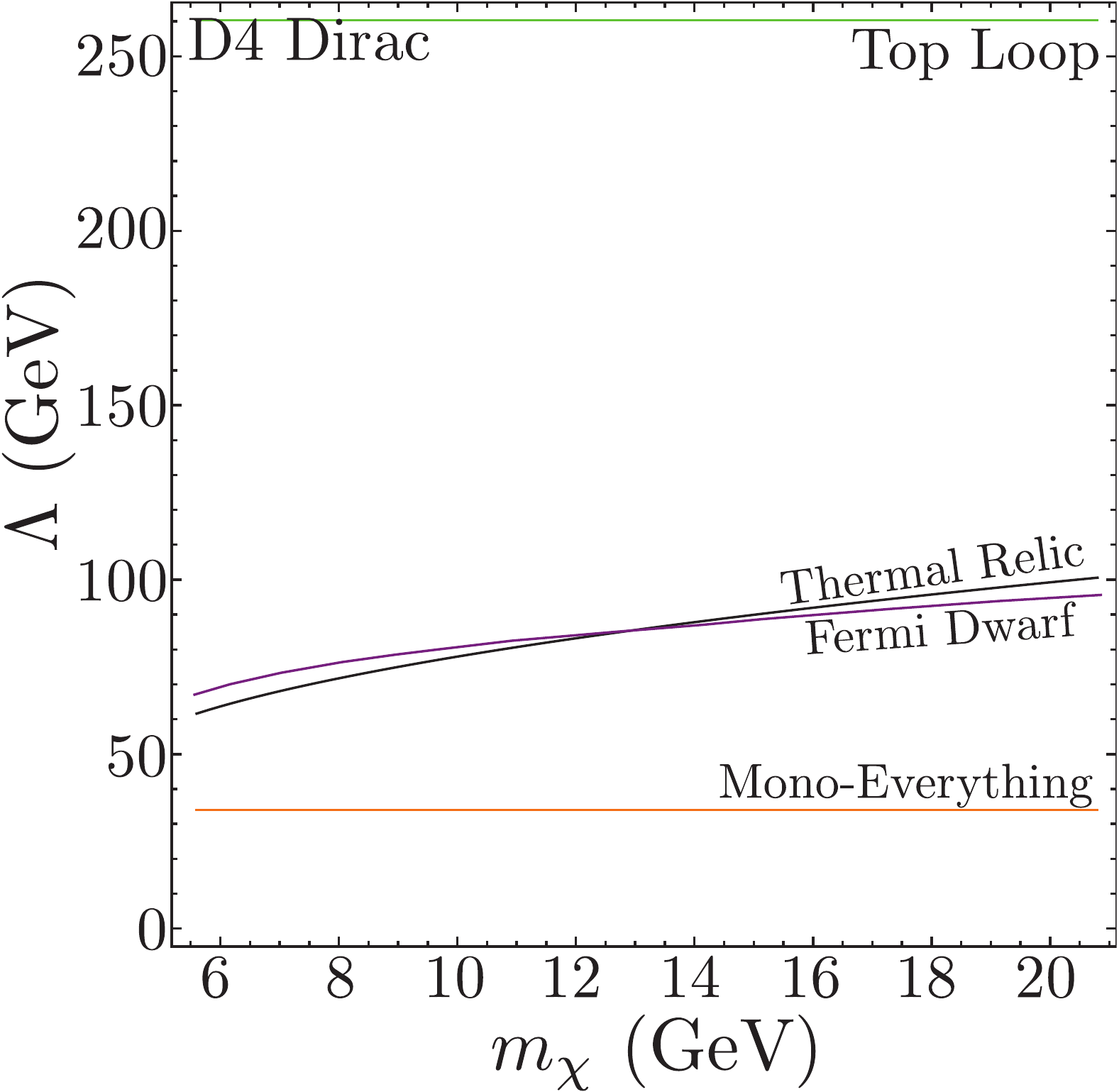}~\includegraphics[width=0.25\columnwidth]{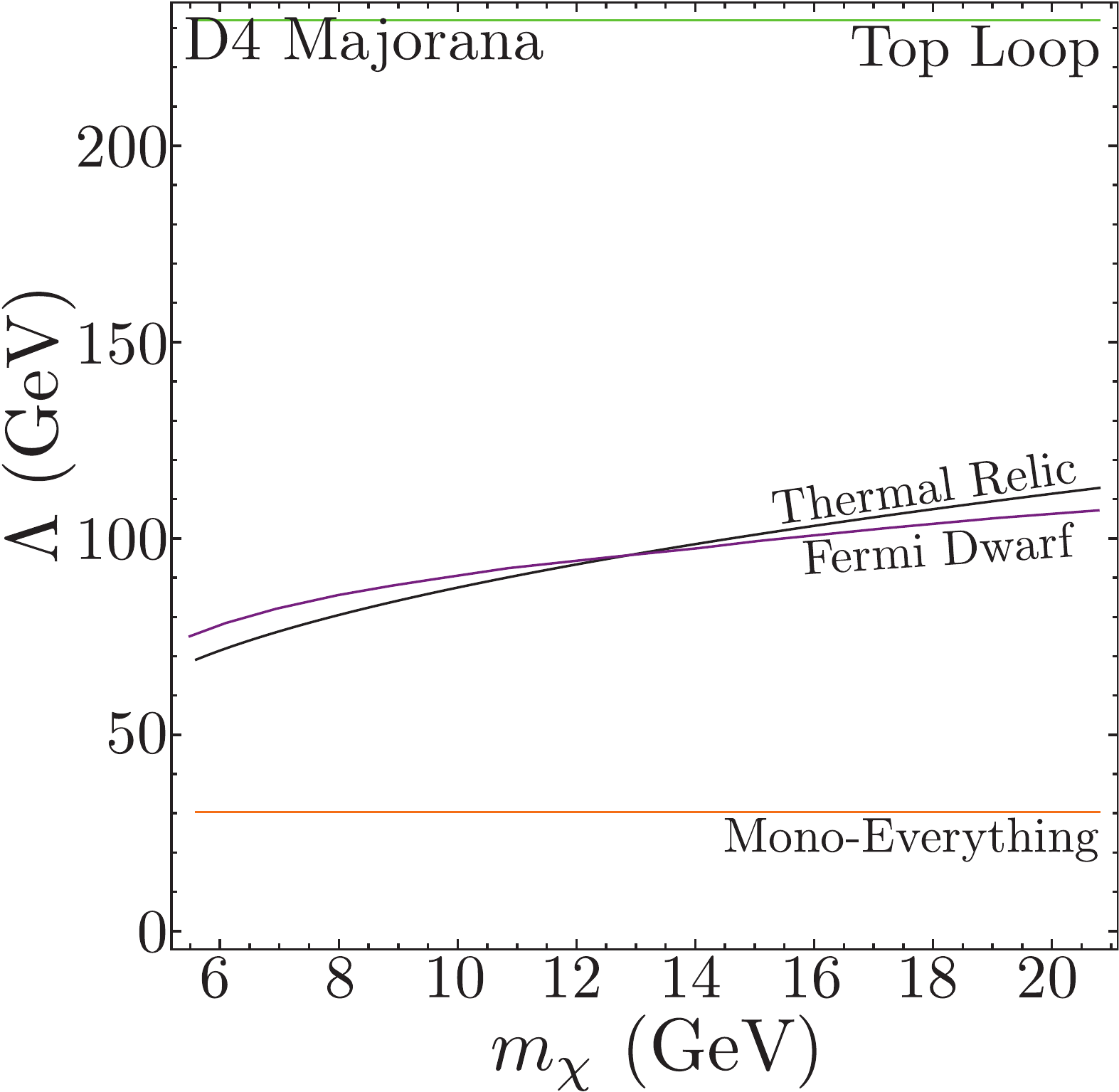}~\includegraphics[width=0.25\columnwidth]{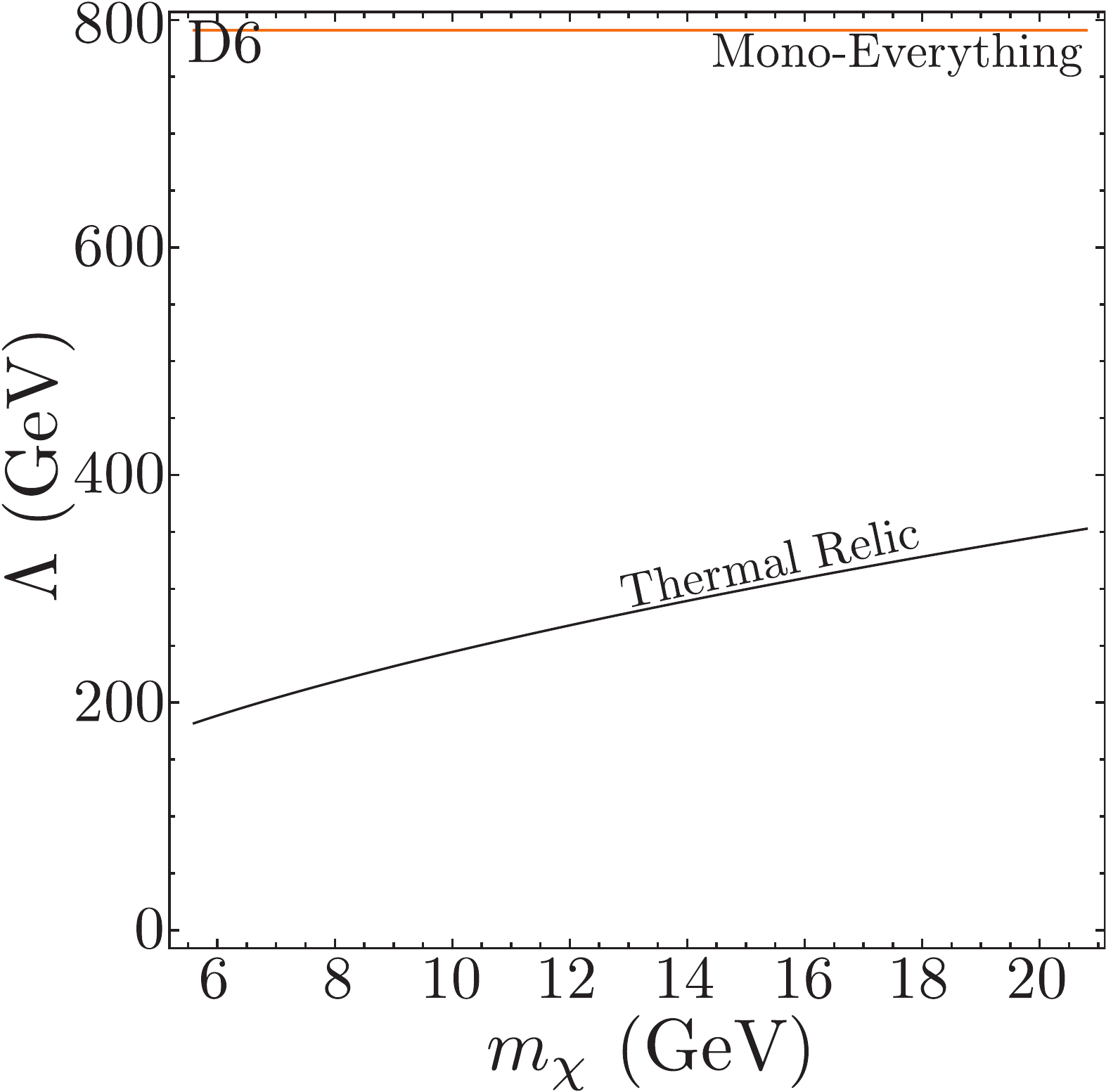}~\includegraphics[width=0.25\columnwidth]{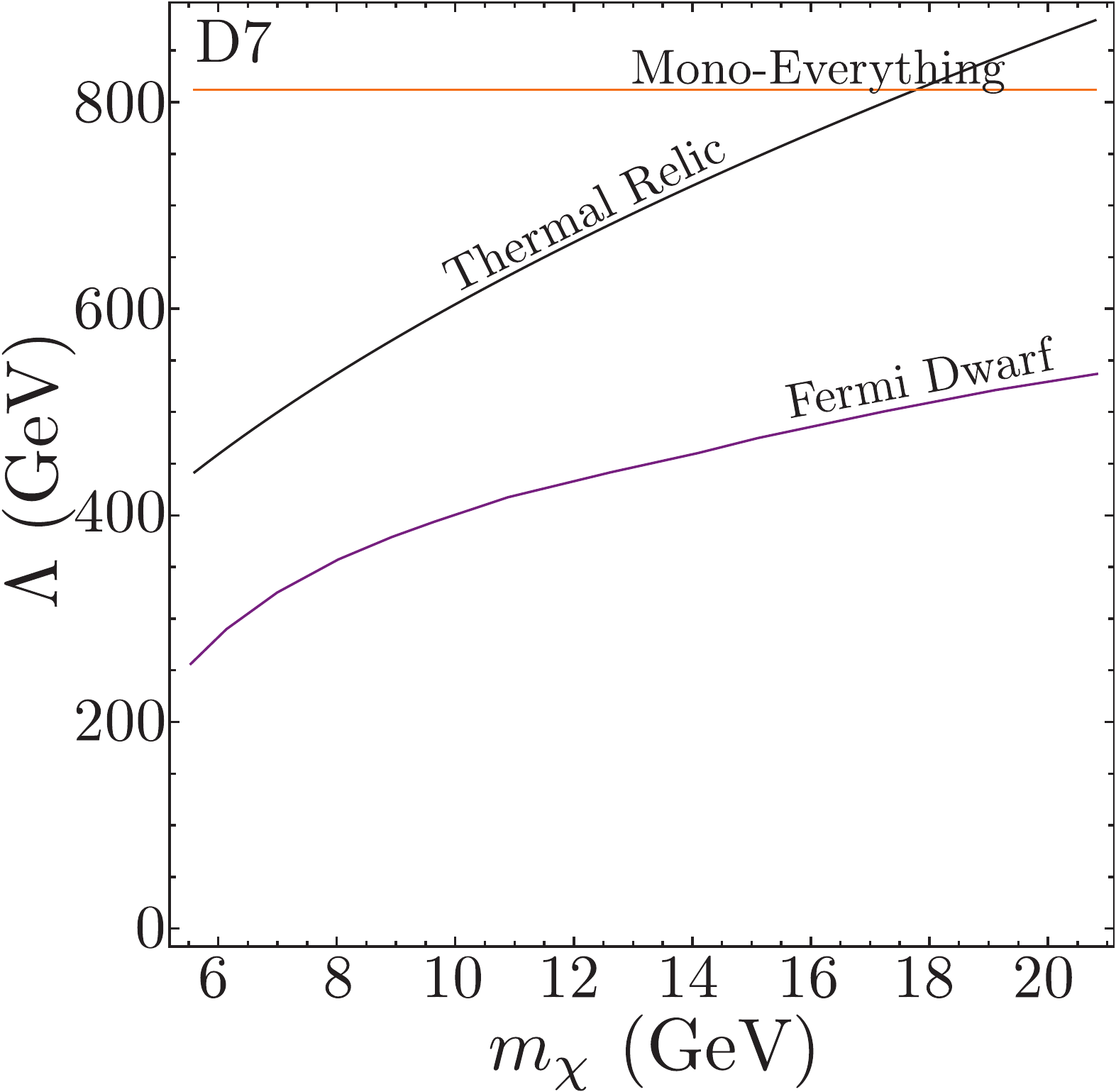}
\includegraphics[width=0.25\columnwidth]{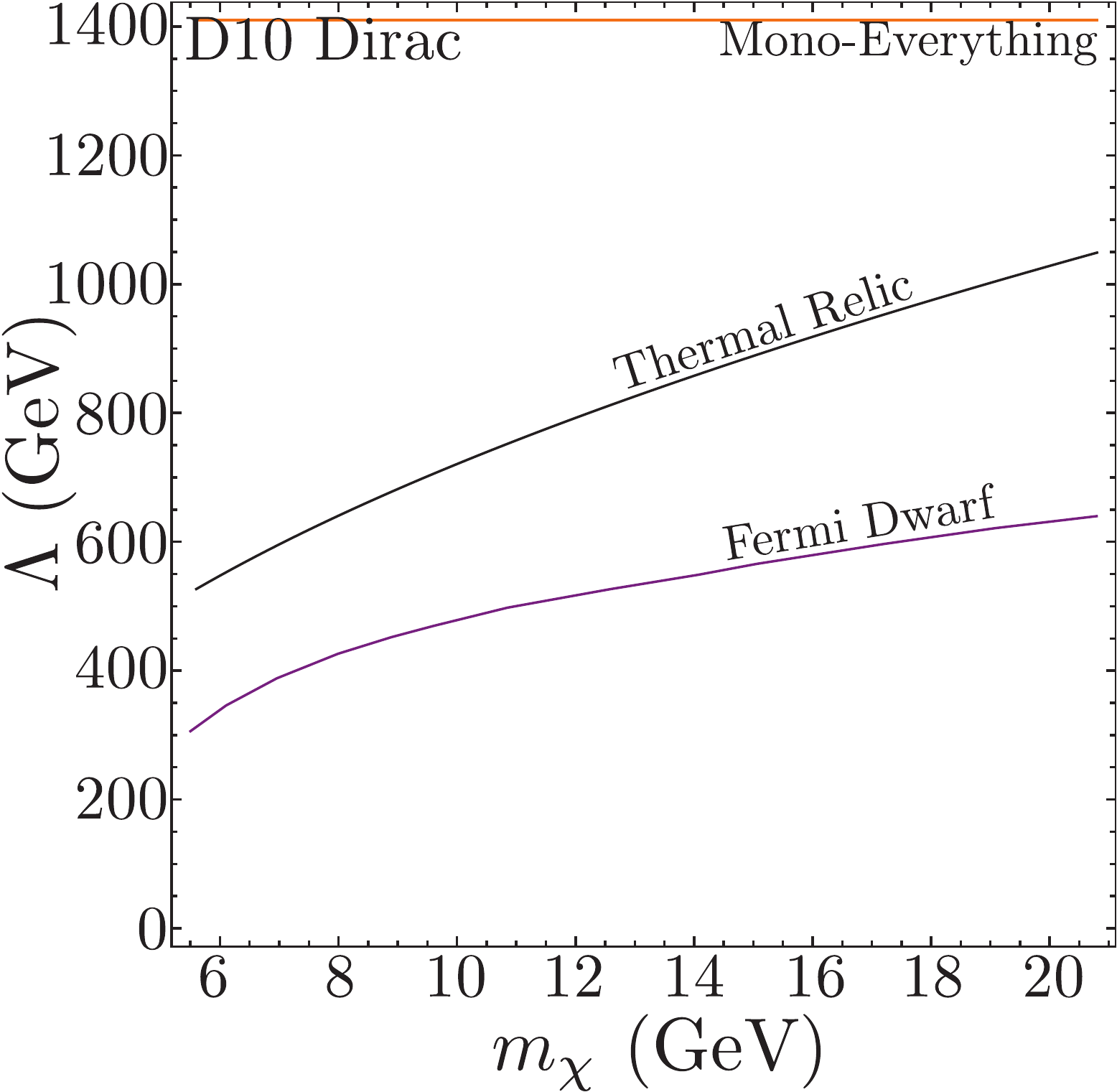}~\includegraphics[width=0.25\columnwidth]{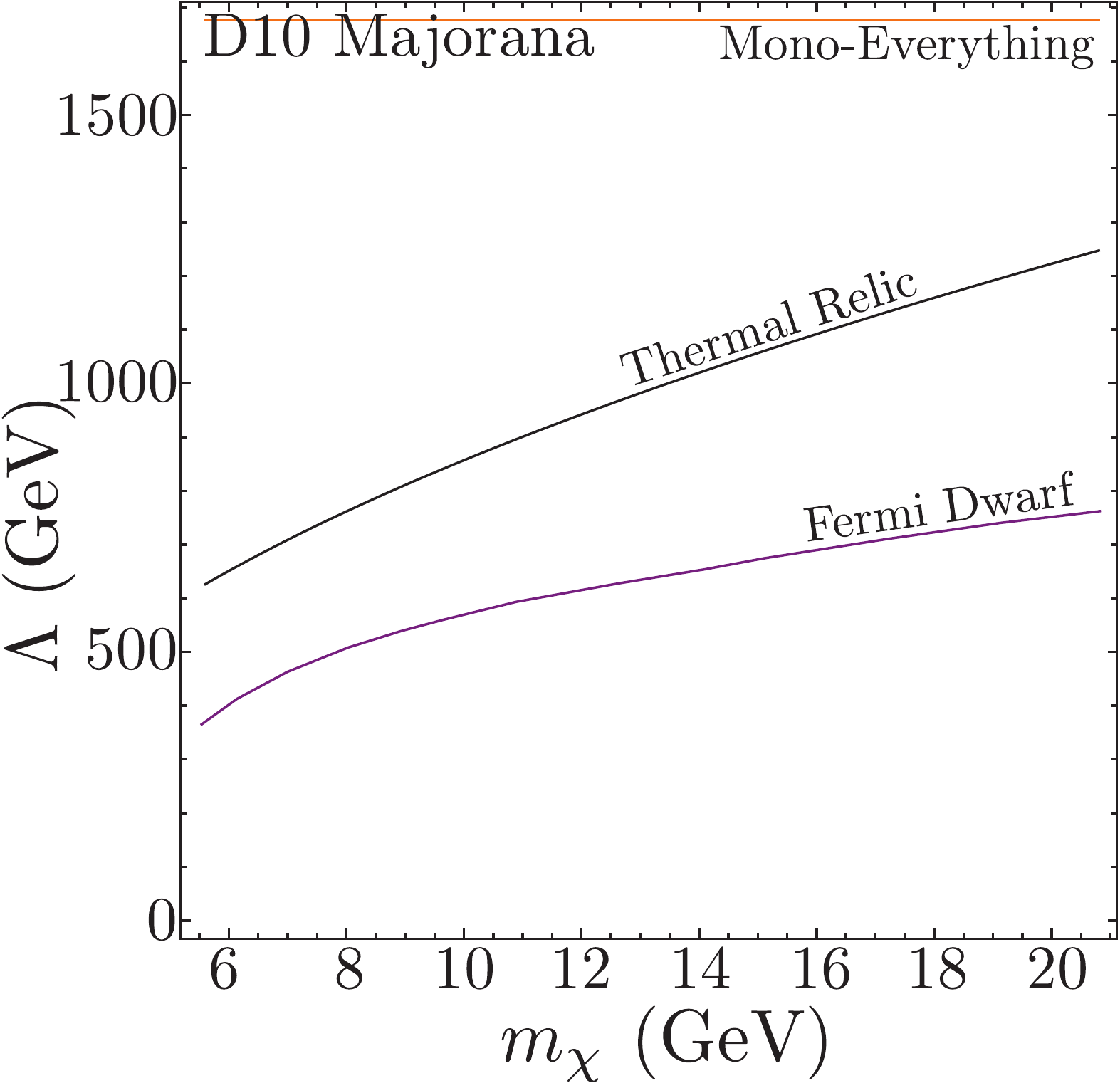}~\includegraphics[width=0.25\columnwidth]{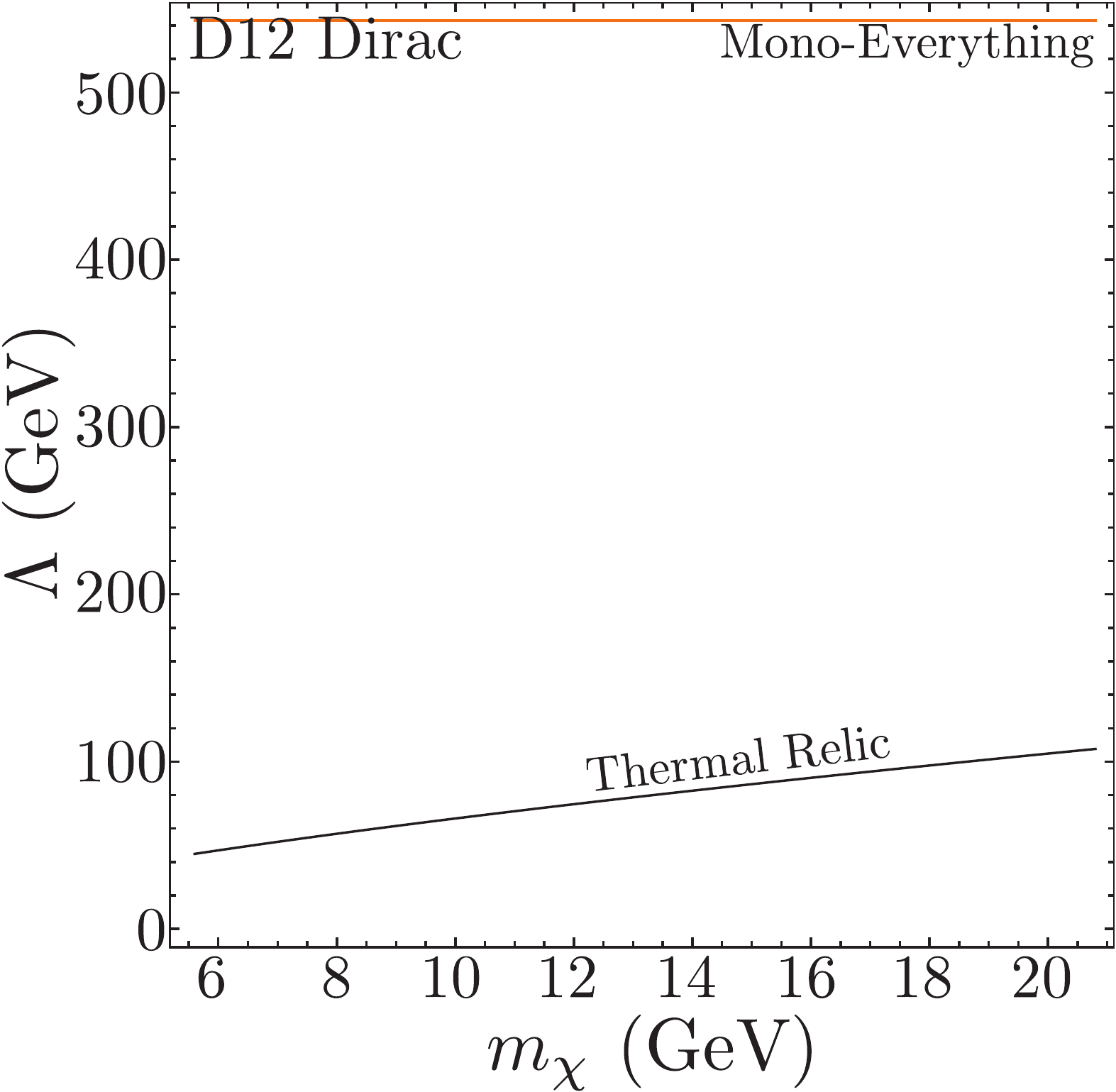}~\includegraphics[width=0.25\columnwidth]{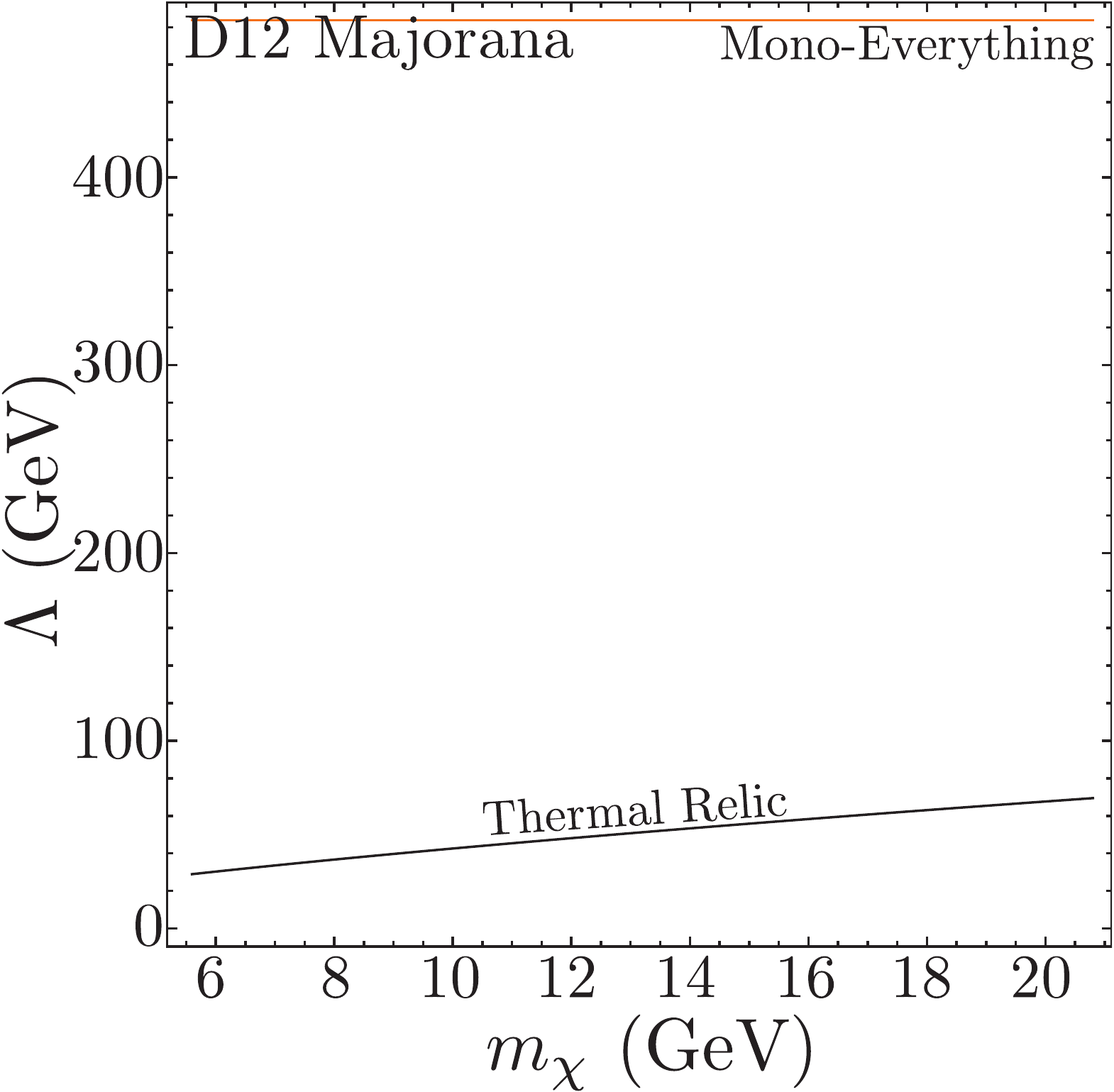}
\includegraphics[width=0.25\columnwidth]{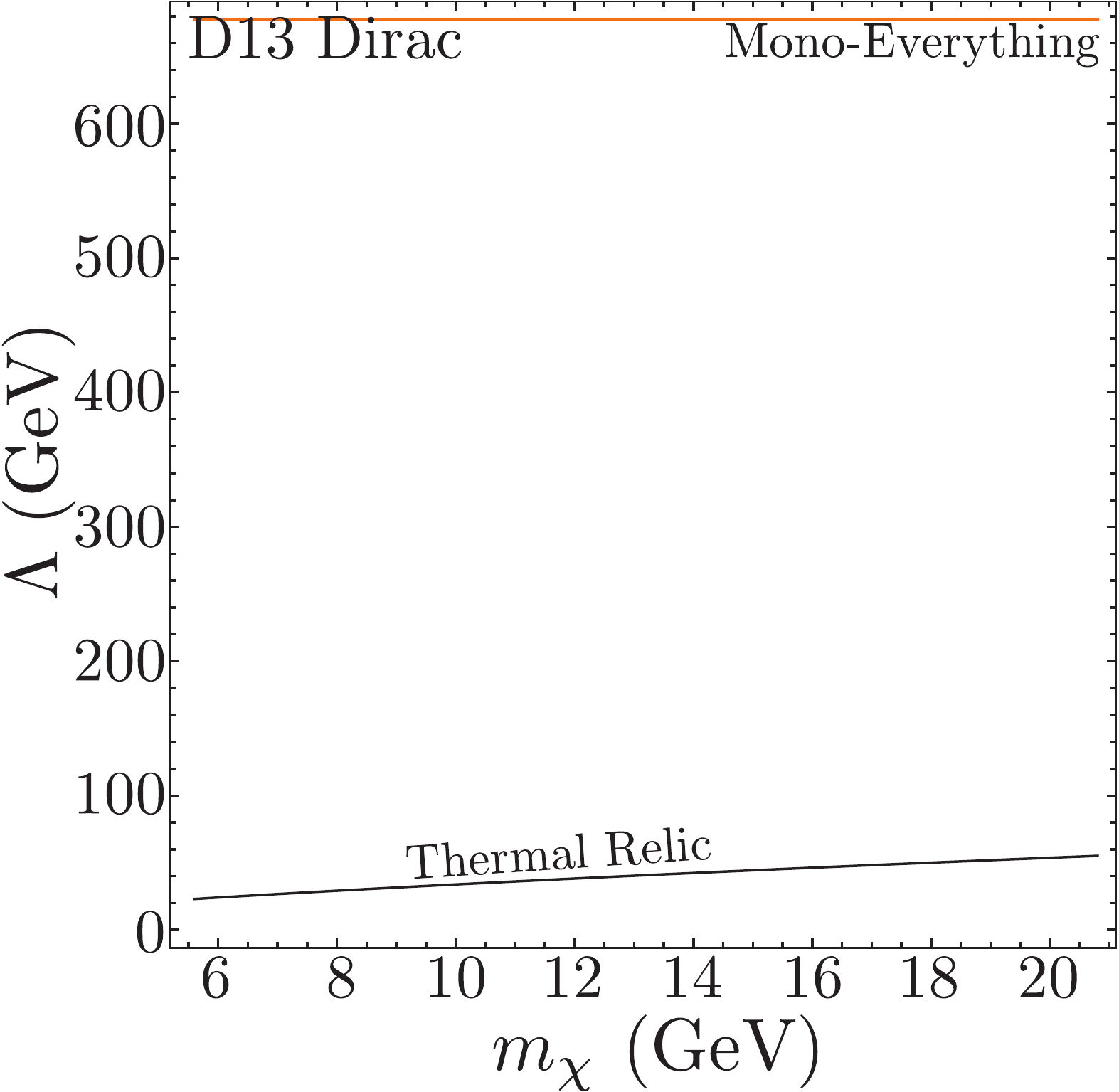}~\includegraphics[width=0.25\columnwidth]{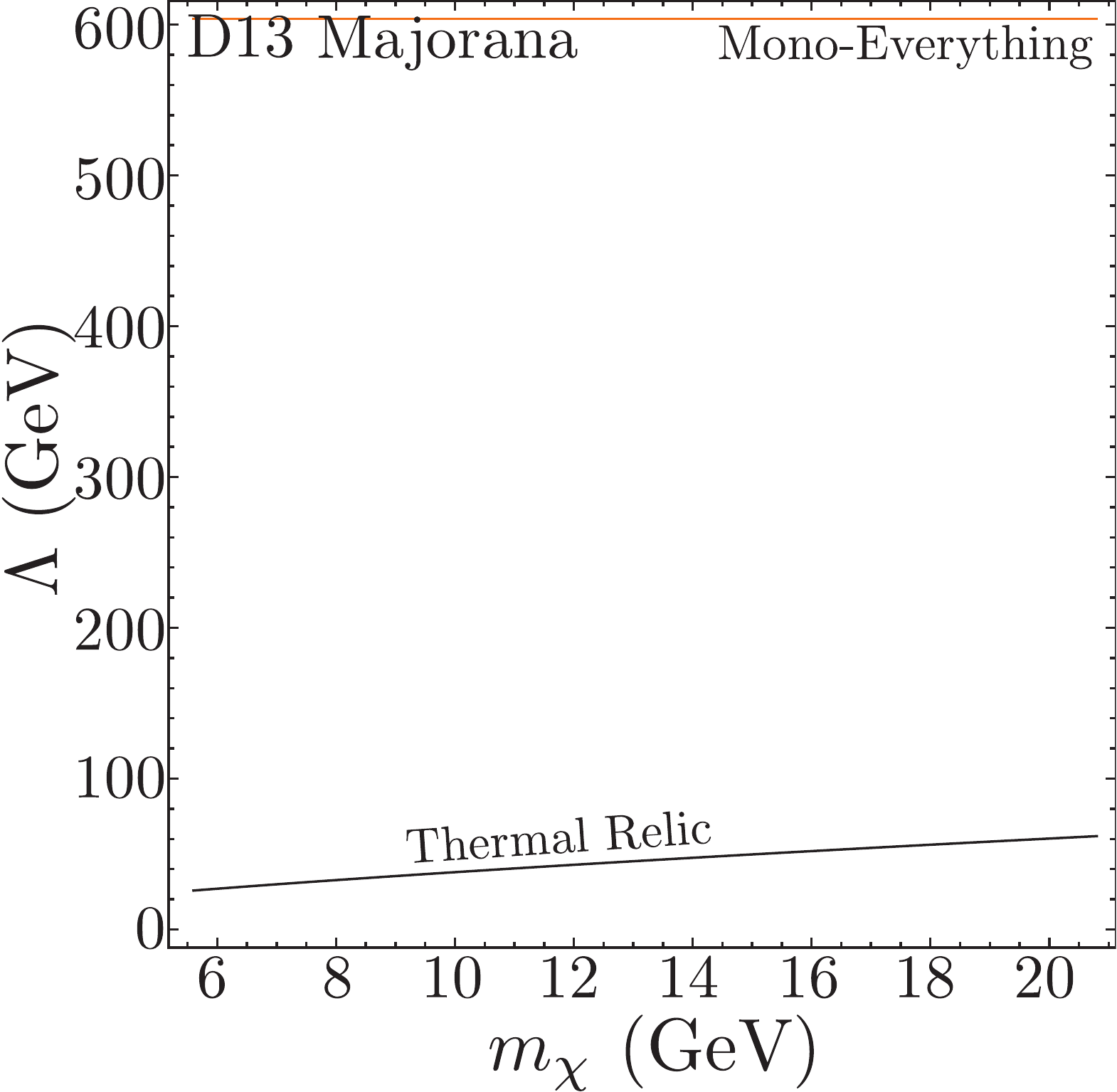}~\includegraphics[width=0.25\columnwidth]{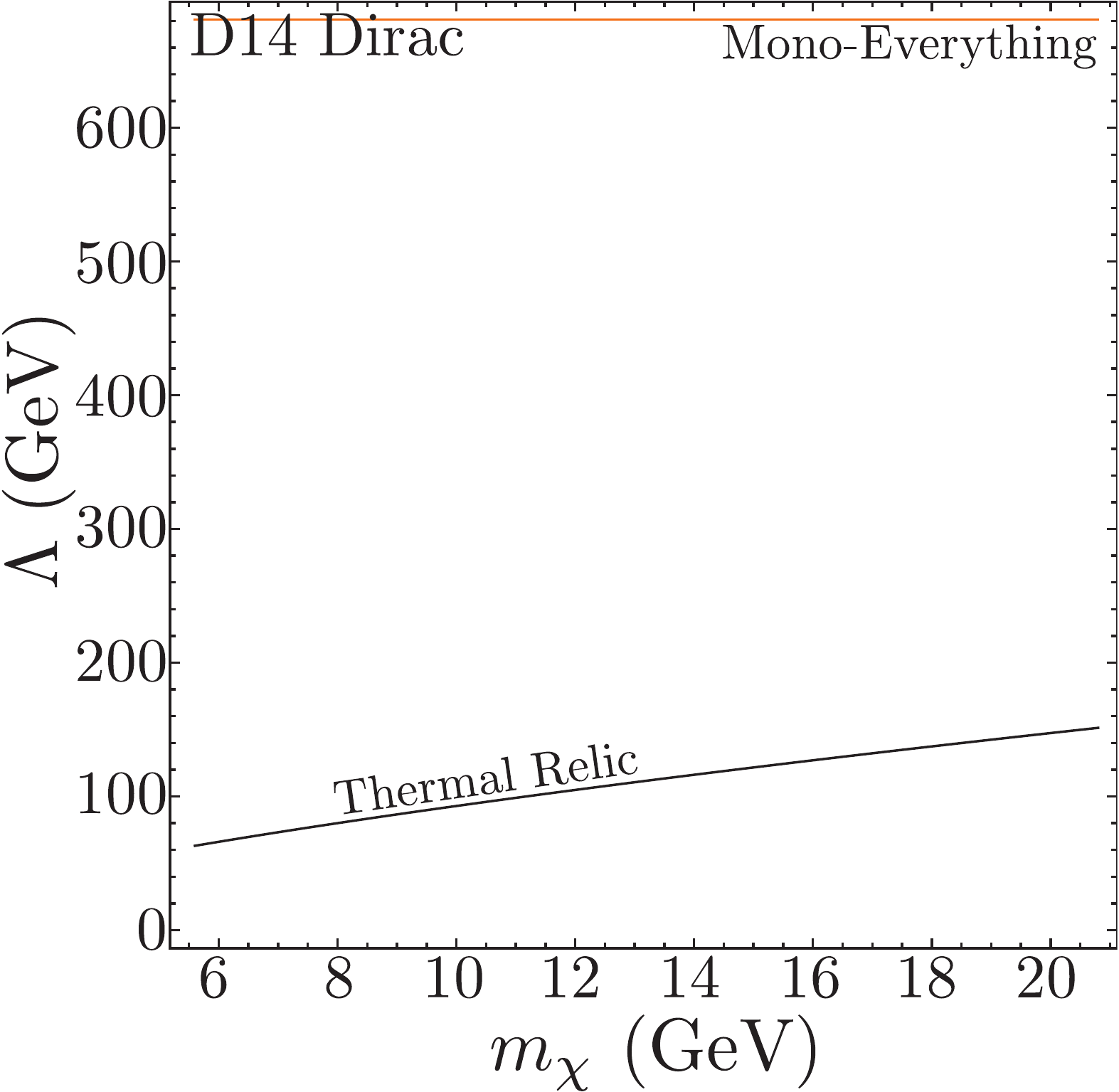}~\includegraphics[width=0.25\columnwidth]{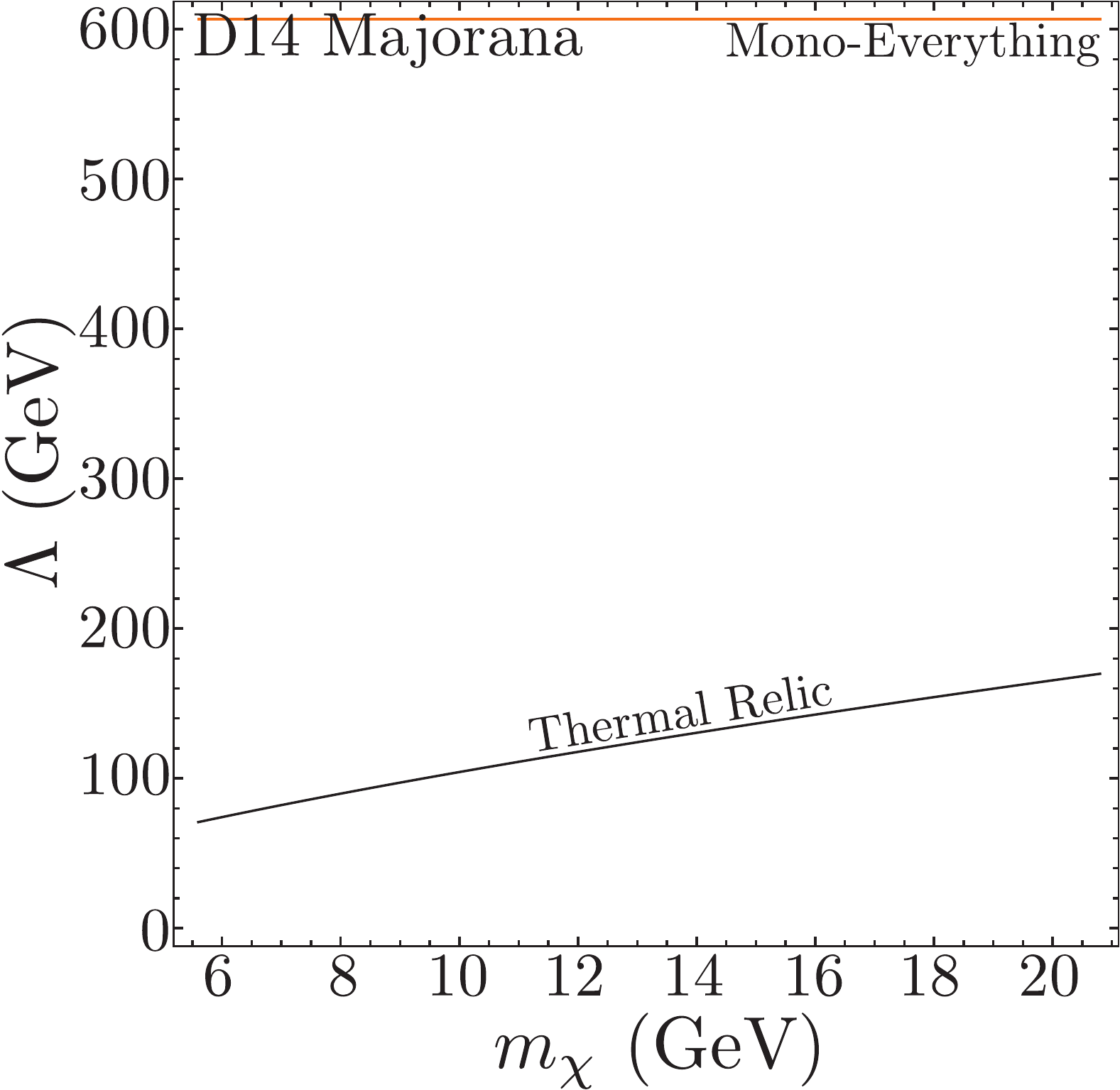}

\caption{Effective operator energy scale $\Lambda$ as a function of $m_\chi$ giving the correct thermal relic abundance of dark matter (black line) and the lower limit on $\Lambda$ from the ``mono-everything'' collider searches \cite{Zhou:2013fla} (orange line).  For operators with couplings proportional to quark mass, the limit of Ref.~\cite{Haisch:2012kf}, derived from the top-loop induced production, is shown as a green line. The operators are those from Table~\ref{tab:listofoperators} that do not give direct detection signals, assuming fermionic dark matter. 
\label{fig:others}}
\end{figure}

\begin{figure}[ht]
\includegraphics[width=0.25\columnwidth]{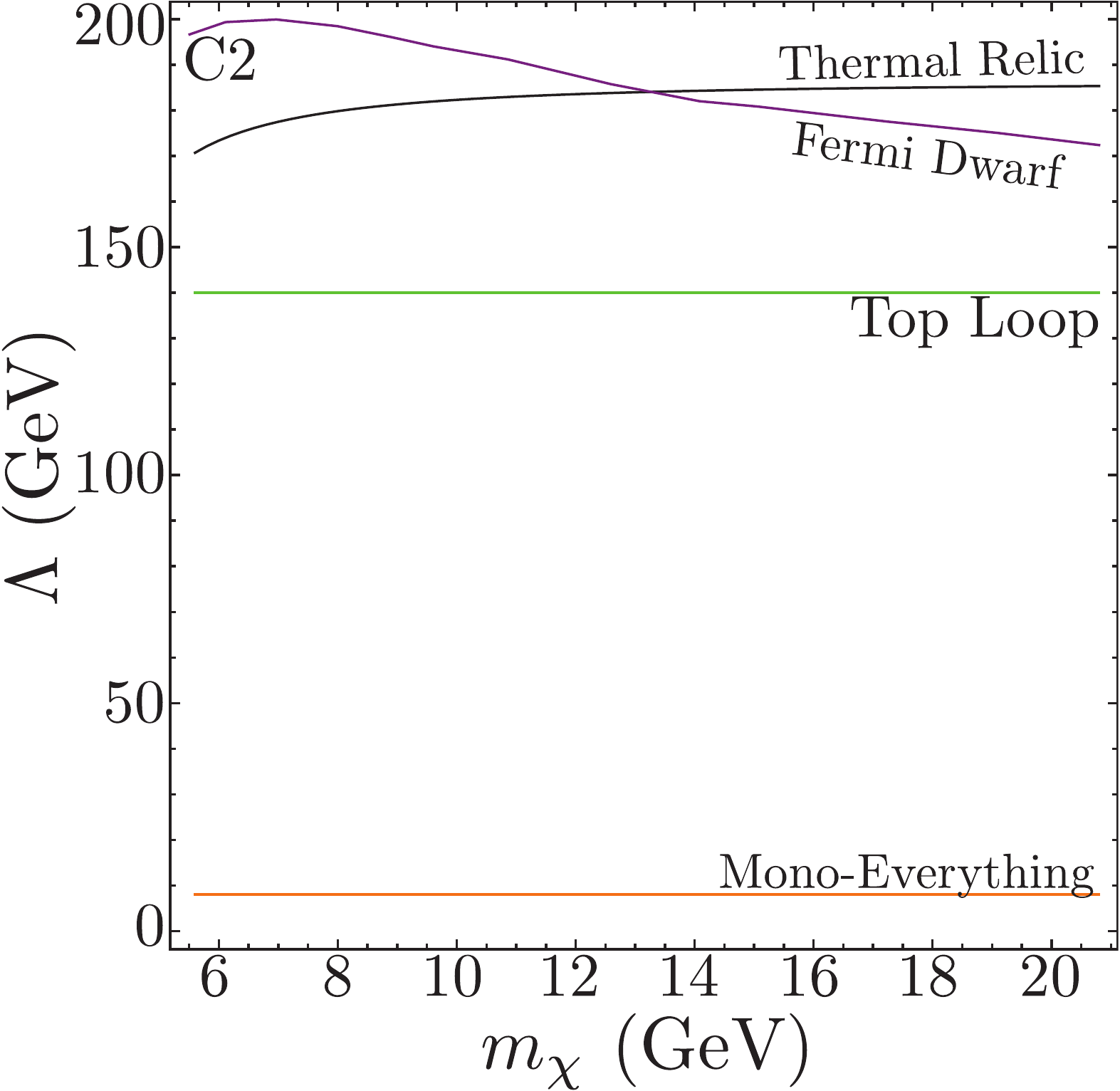}~\includegraphics[width=0.25\columnwidth]{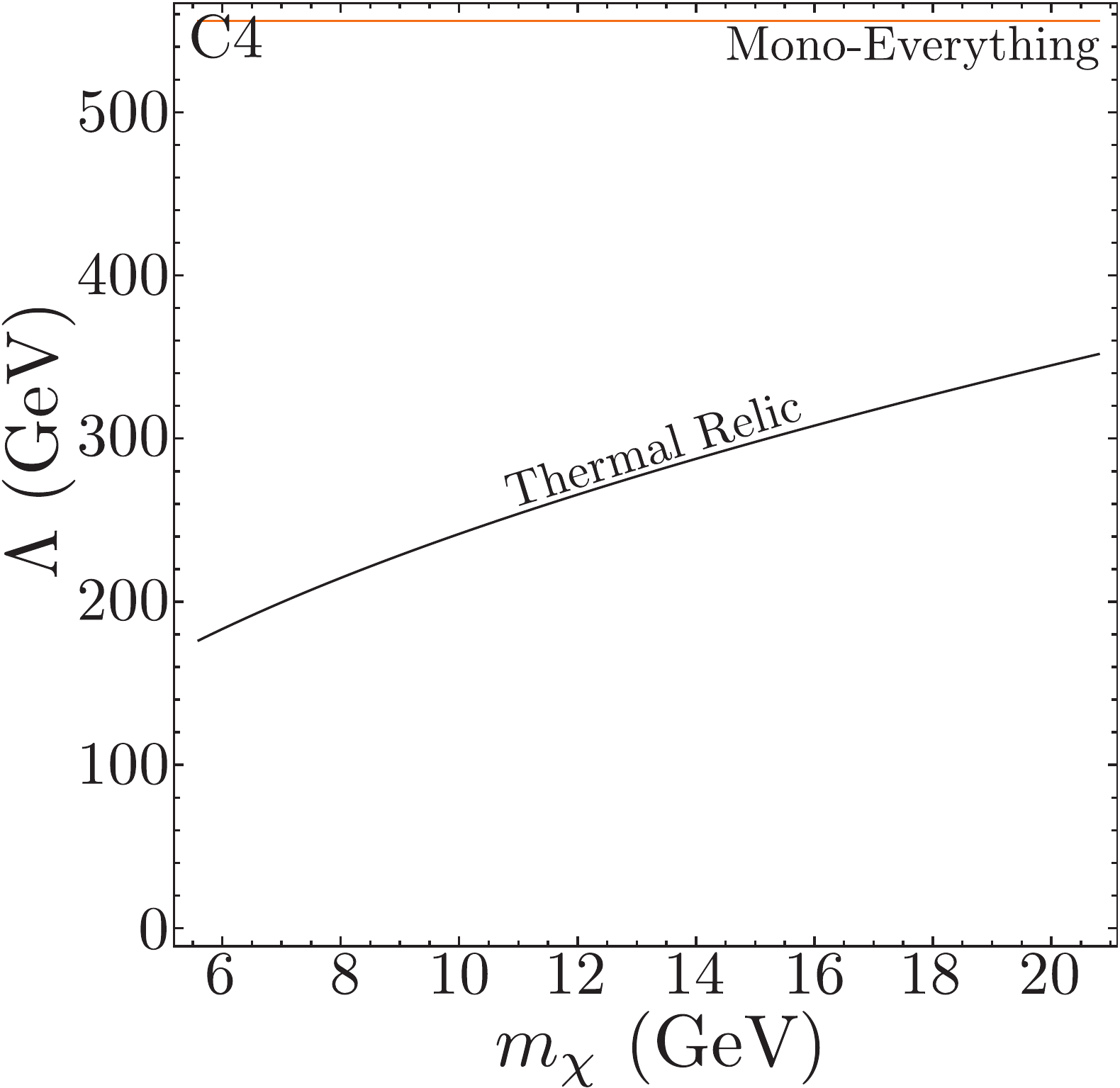}~\includegraphics[width=0.25\columnwidth]{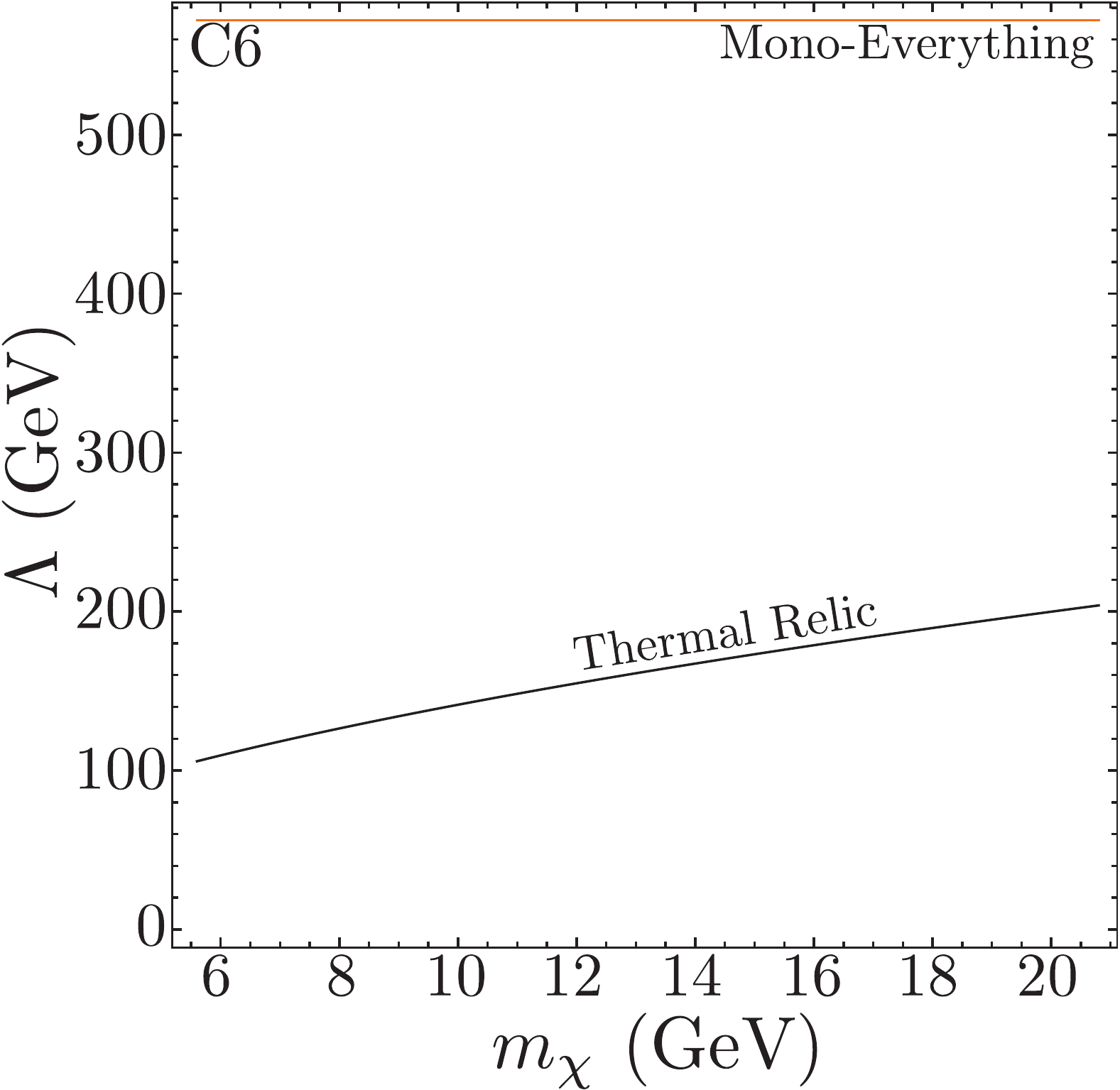}
\includegraphics[width=0.25\columnwidth]{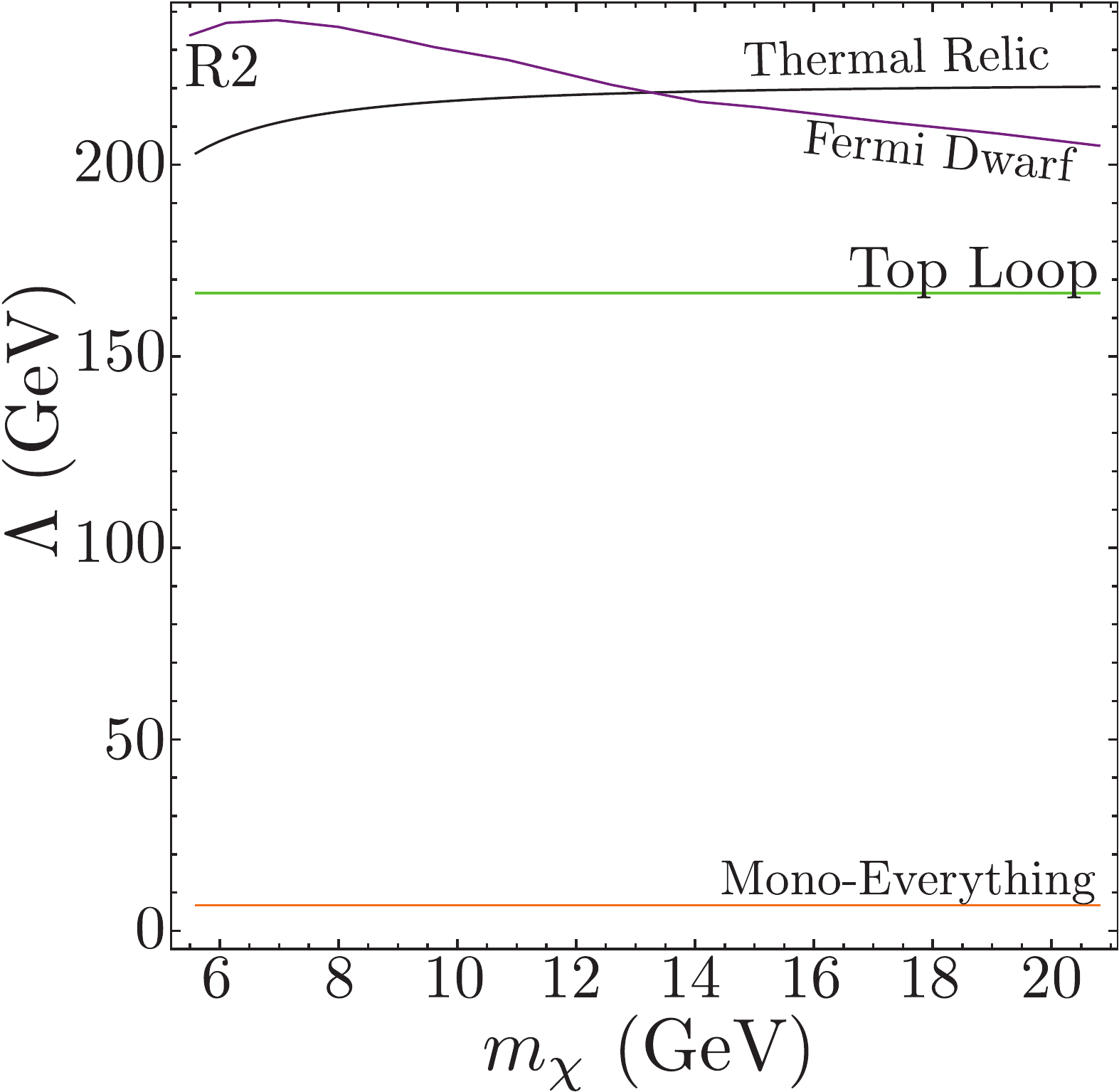}~\includegraphics[width=0.25\columnwidth]{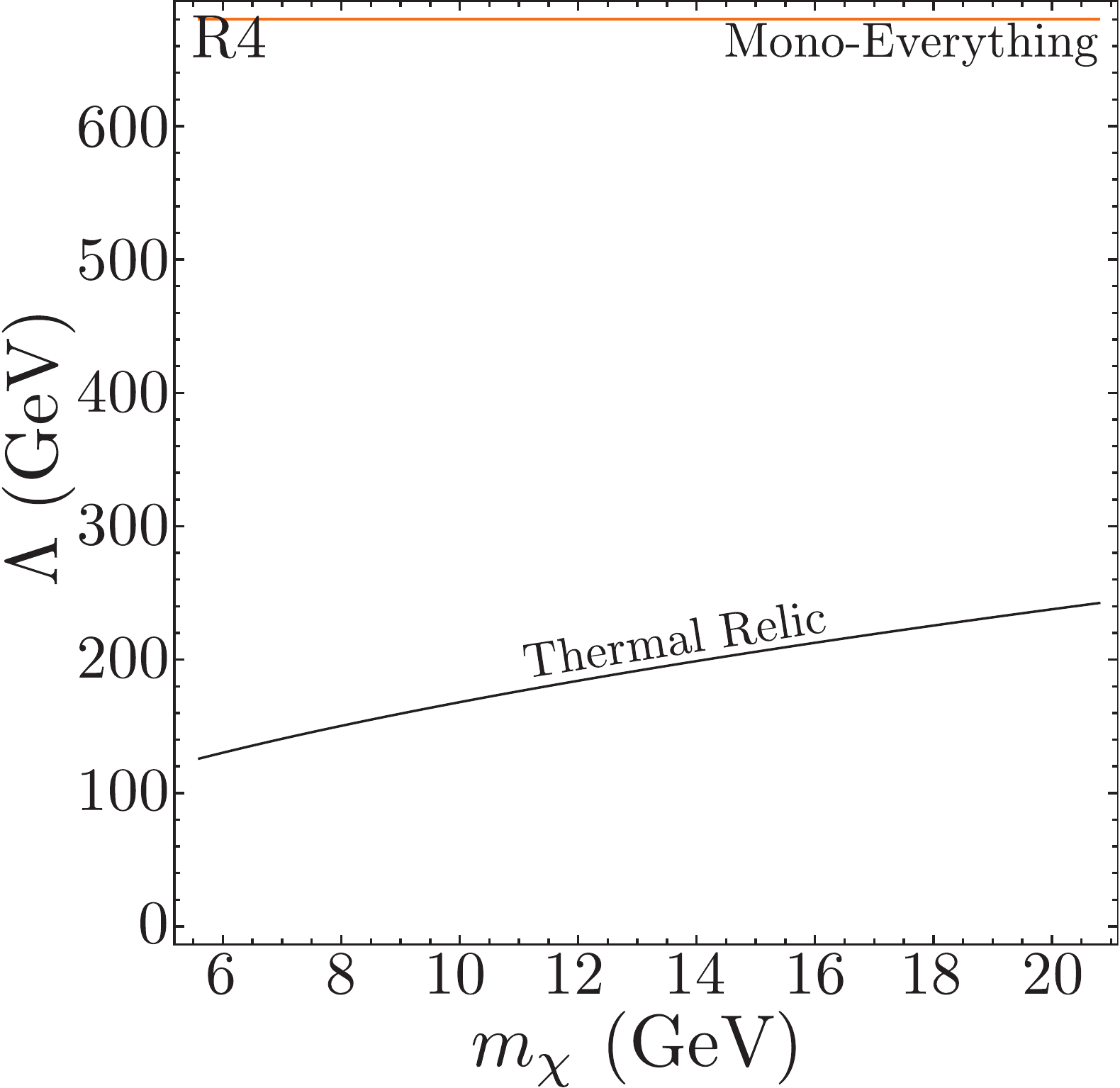}

\caption{Effective operator energy scale $\Lambda$ as a function of $m_\chi$ giving the correct thermal relic abundance of dark matter (black line) and the lower limit on $\Lambda$ from the ``mono-everything'' collider searches \cite{Zhou:2013fla} (orange line). For operators with couplings proportional to quark mass, the limit of Ref.~\cite{Haisch:2012kf}, derived from the top-loop induced production, is shown as a green line. The operators are those from Table~\ref{tab:listofoperators} that do not give direct detection signals, assuming scalar dark matter.  \label{fig:others2}}
\end{figure}

The required thermal relic values and constraints on $\Lambda$ for fermionic dark matter are shown in Fig.~\ref{fig:others}, and for scalar dark matter in Fig.~\ref{fig:others2}. As can be seen, all of the fermionic operators are definitively ruled out by collider constraints. This includes those operators (D2, D3, and D4) which have couplings proportional to Standard Model fermion mass. The values of $\Lambda$ required to obtain the correct relic abundance are not excluded by the straight-forward mono-everything searches, but rather by extrapolation of the induced gluon-coupling that would be produced by the top-loop. Sufficiently low values of $\Lambda$ could be obtained in these cases if the coupling was not flavor-universal. That is, the collider bounds could be evaded if the coupling to the top-quark was significantly lower than that predicted by the minimal-flavor violating assumption. 

Fig.~\ref{fig:others2} shows the values of $\Lambda$ that result in scalar dark matter having the correct thermal abundance, along with the collider constraints. The operators that do not have couplings proportional to fermion mass (C4, C5 and R4) are decisively ruled out by the collider searches. Operators C2 and R2 are not ruled out, similarly to the direct detection-inducing operators C1 and R1. Again, this suggests that the only class of effective operator that explains the CoGeNT/CDMS-Si signal with a thermal relic are ones that couple scalar dark matter through scalar and/or pseudo-scalar operators. The required scales $\Lambda \sim 200$~GeV for thermal dark matter from operators C2 and R2 are not clearly out of the regime of validity for effective operators (due to the small couplings with the light quarks), but closer study is required.  

\section{Conclusion \label{sec:conclusion}}

In this paper, I considered the effective operator formalism in light of the possible positive signals of light ${\cal O}(10~\mbox{GeV})$ dark matter from direct detection experiments. If these signals are indeed of dark matter, and if the interactions of dark matter and Standard Model particles can be accurately and fully described up to some energy scale $\Lambda$ by and effective operator, then the operator or operators must provide three things. First, it must give the correct direct detection cross section at CoGeNT and CDMS-Si. Second, it must give the correct thermal abundance from pair annihilation in the early Universe. Third, it must evade all the negative results from the LHC missing energy + $X$ searches (dubbed ``mono-everything'' {\it a la} Ref.~\cite{Zhou:2013fla}) and/or the top-loop induced signatures. 

As I have demonstrated in this paper, only one set of effective operators appears to pass all three of these requirements (assuming scalar or fermionic dark matter). Of the dimension-5 and 6 operators (Table~\ref{tab:listofoperators}), only the operators connecting scalar dark matter to Standard Model fermions through scalar interactions can simultaneously provide the required cross sections for the relic abundance and direct detection. These operators do not appear to be ruled out by direct collider searches in the mono-everything searches, through top-loop induced gluon-dark matter interactions (see Refs.~\cite{Busoni:2013lha,Profumo:2013hqa}), or by the dwarf galaxy indirect results. However, the required scale is low enough that more careful study of the collider constraints is necessary (see the forthcoming work Ref.~\cite{ZurekPapucci}). The tension with the Fermi results would suggest that either both scalar and pseudo-scalar operators are acting to create a thermal relic, or that heavy quark coupling is suppressed which would require additional theoretical explanation.

The large couplings that these scalar effective operators have with bottom and top quarks suggest that the mono-everything searches are not the best search channel at the LHC. While this set of operators is not in tension with the predicted bounds from $b$- and $t$-enriched channels \cite{Lin:2013sca}, the suggested searches are systematic limited and possible improvements might be possible with better understanding of tops at the LHC. Furthermore, it is not clear whether the effective formalism will apply to dark matter/top couplings (due to the large top mass), which could lead to more spectacular (though model-dependent) signatures at the LHC. If the results from CoGeNT and CDMS-Si are borne out by future work, this analysis gives strong motivation to a dedicated effort to improve searches for dark matter in association with tops.

Of the remaining operators, even relaxing the requirement that a single operator be responsible for both direct detection signals and the relic abundance in the early Universe does not yield a consistent picture that can explain the proposed CoGeNT/CDMS-Si results solely in terms of effective interactions of dark matter. Adding a second effective operator that does not contribute to the direct detection but does allow for a large early Universe annihilation cross section would na\"{i}vely appear to be a solution, but all such operators are directly excluded by collider null results (other than the scalar and pseudo-scalar operators discussed above). 

If dark matter that is suggested at CoGeNT and CDMS-Si is not a scalar coupling through scalar (and possible pseudo-scalar) operators, then this would imply that either:
\begin{itemize}
\item that the flavor structure of the effective operators is non-trivial -- which could potentially alleviate or remove the top-loop induced constraints,

\item that the effective operators responsible for the early Universe annihilation must couple dark matter only to leptons (see, for example Ref.~\cite{Buckley:2013sca}) -- avoiding the quark- and gluon-based collider bounds completely,
\item or that the effective formalism is not applicable at one or more of energy regimes relevant for collider, early Universe, or direct detection. 
\end{itemize}
The break-down of the effective formalism is indicative of the presence of additional particles which are light compared to the relevant energy scales. This could take the form of light mediators responsible for a comparatively long-range force (see {\it e.g.}~\cite{Hooper:2012cw}) or it could be due to a new particle, heavier than the dark matter itself, which could be produced directly at colliders. As the constraints on new colored particles from the LHC are very stringent this would also seem to imply that the most likely breakdown of the effective theory is the result of some new colorless particle coupling dark matter to the Standard Model. Generically, this would imply that dark matter should have large annihilation channels into leptons or electroweak gauge bosons.

Each of these possible results would be extremely interesting, implying new physics at relatively low energy scales, or that the new physics of dark matter might shed light on the flavor puzzle. If the results of CoGeNT and CDMS survive the necessary experimental cross checks and are borne out by other direct detection results, then this would imply that additional new physics should be accessible in the current or near-future experimental programs.

\section*{Acknowledgements}

I would like to thank Dan Hooper, Ulrich Haisch, and Kathryn Zurek for useful discussions and advice. Fermilab is operated by Fermi Research Alliance, LLC, under contract DE-AC02-07CH11359 with the United States Department of Energy.

\appendix

\section{Thermal Cross Sections \label{app:thermal}}

Here I report the thermally averaged dark matter annihilation cross sections for the effective operators in Table~\ref{tab:listofoperators}. The sums run over fermion species $f$, and $N_f$ is the color factor associated with the fermions ($N_f =3$ for quarks, $1$ for leptons). These results compare (to ${\cal O}(m_f^2)$ in the ${\cal O}(T)$ term) to those found in Ref.~\cite{MarchRussell:2012hi}.
\begin{eqnarray}
\langle \sigma v\rangle^{\rm D1} & = &  \frac{1}{\Lambda^6} \sum_f N_f \frac{3m_f^2(m_\chi^2-m_f^2)^{3/2}}{4\pi m_\chi^2 }T, \\
\langle \sigma v\rangle^{\rm D2} & = &  \frac{1}{\Lambda^6} \sum_f N_f \left(\frac{m_f^2(m_\chi^2-m_f^2)^{3/2}}{2\pi m_\chi }+ \frac{9m_f^4(m_\chi^2-m_f^2)}{8\pi m_\chi^3 }T\right), \\
\langle \sigma v\rangle^{\rm D3} & = &  \frac{1}{\Lambda^6}\sum_f N_f \frac{3m_\chi m_f^2 \sqrt{1-\frac{m_f^2}{m_\chi^2}}}{4\pi }T,\\
\langle \sigma v\rangle^{\rm D4} & = &  \frac{1}{\Lambda^6} \sum_f N_f \left(\frac{m_\chi^2m_f^2 \sqrt{1-\frac{m_f^2}{m_\chi^2}}}{2\pi}+ \frac{3m_f^4}{8\pi m_\chi \sqrt{1-\frac{m_f^2}{m_\chi^2}}}T\right),\\
\langle \sigma v\rangle^{\rm D5} & = &  \frac{1}{\Lambda^4}\sum_f N_f \left(\frac{(2m_\chi^2+m_f^2)\sqrt{1-\frac{m_f^2}{m_\chi^2}}}{2\pi} + \frac{11m_f^2+2m_f^2m_\chi^2-4m_\chi^4}{8\pi m_\chi^3\sqrt{1-\frac{m_f^2}{m_\chi^2}} }T\right), \\
\langle \sigma v\rangle^{\rm D6} & = & \frac{1}{\Lambda^4}\sum_f N_f \frac{(2m_\chi^2+m_f^2)\sqrt{1-\frac{m_f^2}{m_\chi^2}}}{2\pi m_\chi }T, \\
\langle \sigma v\rangle^{\rm D7} & = & \frac{1}{\Lambda^4}\sum_f N_f \left(\frac{(m_\chi^2-m_f^2)\sqrt{1-\frac{m_f^2}{m_\chi^2}}}{\pi }-\frac{(2m_\chi^4-13 m_\chi^2m_f^2+11m_f^4)}{2\pi m_\chi^3\sqrt{1-\frac{m_f^2}{m_\chi^2}}}T \right), \\
\langle \sigma v\rangle^{\rm D8} & = & \frac{1}{\Lambda^4}\sum_f N_f\left(\frac{m_f^2\sqrt{1-\frac{m_f^2}{m_\chi^2}}}{2\pi}+\frac{23m_f^4-28m_f^2m_\chi^2+m_\chi^4}{8\pi m_\chi^3  \sqrt{1-\frac{m_f^2}{m_\chi^2}}}T \right),\\
\langle \sigma v\rangle^{\rm D9} & = & \frac{1}{\Lambda^4}\sum_f N_f\left(\frac{2(m_\chi^2+2m_f^2)\sqrt{1-\frac{m_f^2}{m_\chi^2}}}{\pi} + \frac{28m_f^4-17m_f^2m_\chi^2-2m_\chi^4}{2\pi m_\chi^3\sqrt{1-\frac{m_f^2}{m_\chi^2}}}T\right), \\
\langle \sigma v\rangle^{\rm D10} & = & \frac{1}{\Lambda^4}\sum_f N_f\left(\frac{2(m_\chi^2-m_f^2)\sqrt{1-\frac{m_f^2}{m_\chi^2}}}{\pi}-\frac{(2m_\chi^2-19m_f^2m_\chi^2+17m_f^4}{2\pi m_\chi^3 \sqrt{1-\frac{m_f^2}{m_\chi^2}}}T\right)
\end{eqnarray}
\begin{eqnarray}
\langle \sigma v\rangle^{\rm D11} & = & \frac{\alpha_S^2}{\Lambda^6}\left( \frac{12 m_\chi^3}{\pi} T \right), \\
\langle \sigma v\rangle^{\rm D12} & = &  \frac{\alpha_S^2}{\Lambda^6}\left( \frac{8 m_\chi^4}{\pi}+\frac{12 m_\chi^3}{\pi}T \right),  \\
\langle \sigma v\rangle^{\rm D13} & = & \frac{\alpha_S^2}{\Lambda^6}\left( \frac{6 m_\chi^3}{\pi} T \right), \\
\langle \sigma v\rangle^{\rm D14} & = &  \frac{\alpha_S^2}{\Lambda^6}\left( \frac{64 m_\chi^4}{\pi}+\frac{6 m_\chi^3}{\pi}T \right), \\
\langle \sigma v\rangle^{\rm C1/R1} & = &\frac{1}{\Lambda^4}  \sum_f N_f\left(\frac{m_f^2(m_\chi^2-m_f^2)^{3/2}}{4\pi m_\chi^3}-\frac{3m_f^2(2m_\chi^4-7m_f^2m_\chi^2+5m_f^4)}{16\pi m_\chi^5\sqrt{1-\frac{m_f^2}{m_\chi^2}}} T \right), \\
\langle \sigma v\rangle^{\rm C2/R2} & = &  \frac{1}{\Lambda^4} \sum_f N_f \left(\frac{m_f^2 \sqrt{1-\frac{m_f^2}{m_\chi^2}}}{4\pi}-\frac{m_f^2(2m_\chi^2-3m_f^2)}{16\pi m_\chi^3 \sqrt{1-\frac{m_f^2}{m_\chi^2}}} T \right)\\
\langle \sigma v\rangle^{\rm C3} & = & \frac{1}{\Lambda^4} \sum_f N_f \frac{(2m_\chi^2+m_f^2)\sqrt{1-\frac{m_f^2}{m_\chi^2}}}{2\pi m_\chi }T, \\
\langle \sigma v\rangle^{\rm C4} & = & \frac{1}{\Lambda^4} \sum_f N_f \frac{(m_\chi^2-m_f^2)\sqrt{1-\frac{m_f^2}{m_\chi^2}}}{\pi m_\chi }T \\
\langle \sigma v\rangle^{\rm C5/R3} & = & \frac{\alpha_S^2}{\Lambda^4}\left(  \frac{2 m_\chi^2}{\pi} \right), \\
\langle \sigma v\rangle^{\rm C6/R4} & = & \frac{\alpha_S^2}{\Lambda^4}\left( \frac{4 m_\chi^2}{\pi} \right).
\end{eqnarray}

\bibliographystyle{apsrev}
\bibliography{cogentdm}
\end{document}